\documentclass[
aps,
physrev,
amsmath,
amssymb,
reprint,
]{revtex4-1}

\usepackage[utf8]{inputenc}
\usepackage{graphicx}
\usepackage[caption=false]{subfig}
\usepackage{braket}
\usepackage{natbib}
\usepackage{color}
\usepackage{xcolor}

\newcommand{\im}{\mathrm{Im}\:}

\newcommand{\nmom}{n_\mathrm{mom}}
\newcommand{\naux}{n_\mathrm{aux}}
\newcommand{\haux}{h_\mathrm{aux}}

\newcommand{\remove}[1]{}

\newcommand{\new}[1]{#1}

\begin{document}

\title{Fully Algebraic and Self-consistent Effective Dynamics in a Static Quantum Embedding}

\author{P.~V.~Sriluckshmy}
\author{Max~Nusspickel}
\author{Edoardo Fertitta}
\author{George~H.~Booth}
\email{george.booth@kcl.ac.uk}
\affiliation{Department of Physics, King's College London, Strand, London, WC2R 2LS, U.K.}

\date{\today}
\begin{abstract}
Quantum embedding approaches involve the self-consistent optimization of a local fragment of a strongly correlated system, entangled with the wider environment. The `energy-weighted' density matrix embedding theory (EwDMET) was established recently as a way to systematically control the resolution of the fragment--environment coupling, and allow for true quantum fluctuations over this boundary to be self-consistently optimized within a fully static framework.
In this work, we reformulate the algorithm to ensure that EwDMET can be considered equivalent to an optimal and rigorous truncation of the self-consistent dynamics of dynamical mean-field theory (DMFT).
A practical limitation of these quantum embedding approaches is often a numerical fitting of a self-consistent object defining the quantum effects. 
However, we show here that in this formulation, all numerical fitting steps can be entirely circumvented, via an effective Dyson equation in the space of truncated dynamics. This provides a robust and analytic self-consistency for the method, and an ability to systematically and rigorously converge to DMFT from a static, wave function perspective. We demonstrate that this improved approach can solve the correlated dynamics and phase transitions of the Bethe lattice Hubbard model in infinite dimensions, as well as one- and two-dimensional Hubbard models where we clearly show the benefits of this rapidly convergent basis for correlation-driven fluctuations. This systematically truncated description of the effective dynamics of the problem also allows access to quantities such as Fermi liquid parameters and renormalized dynamics, and demonstrates a numerically efficient, systematic convergence to the zero-temperature dynamical mean-field theory limit.
\end{abstract}
\maketitle

\section{Introduction}

The competition between strong electronic interactions within local, atomic energy levels with the hybridization to a wider extended bandstructure gives rise to some of the most remarkable emergent properties of quantum matter. These include a variety of quantum phase transitions \cite{RevModPhys.70.1039}, colossal responses to external stimuli \cite{PhysRevLett.123.047201}, electronically-driven superconductivity \cite{PhysRevB.59.6430}, and many other novel states of matter \cite{Wang_2011}. Much of this physics can be qualitatively described via the paradigmatic Hubbard model \cite{doi:10.1098/rspa.1963.0204,wakoh1970}. However, in recent years the merging of the Hubbard interactions with an {\it ab initio} derived bandstructure has led to approaches such as LDA+$U$ and LDA+DMFT, which have become the dominant approach to a qualitatively correct treatment of strongly correlated materials \cite{PhysRevB.57.6884,RevModPhys.78.865,PhysRevLett.86.5345}. While these methods differ in their treatment of the local interaction term, they both nevertheless explicitly and self-consistently consider the effects of this local correlated physics embedded in the mean-field-derived bandstructure. 

\new{In this work, we update and improve upon an emerging and alternative quantum embedding framework, which we term `Energy-weighted density matrix embedding theory' (EwDMET), which allows for these local interactions and wider bandstructure to be consistently coupled. The method now exhibits a number of appealing features, including an improved robust and analytic self-consistency formulation, and a rigorous interpolation between the most widely used quantum embedding theories in the community. We initially review these other methods in a general and combined way, to provide the context to best understand the EwDMET approach through their successes and limitations.}

Dynamical mean-field theory (DMFT) relies on a self-consistent optimization of the local (`fragment') electronic propagator (Green's function), matching it between the extended lattice picture and a local auxiliary (`impurity') model with explicit interactions \cite{georges1996,hettler2000,kotliar2001,senechal2008}. However, the computational overheads which arise when working with propagators and their associated continuous energy or time variables (especially at low or zero temperature), prompted the development of alternatives, where the self-consistent optimization was performed with respect to a different quantum variable~\cite{doi:10.1021/acs.accounts.6b00356}.
Density matrix embedding theory~(DMET) is instead formulated as a self-consistent optimization of the static, fragment-local, one-particle reduced density matrix (RDM) \cite{knizia2012,doi:10.1021/acs.jctc.6b00316}, in a similar spirit to some other self-consistent, static embedding methods \cite{lechermann2007,PhysRevB.96.195126,lee2019}.
This results in significant simplifications compared to DMFT, allowing for larger local fragments to be treated, as well as an algebraic construction of the local auxiliary Hamiltonian.
This efficient formulation has been shown to capture much of the strongly correlated, local physics for both Hubbard and {\it ab initio} models, resulting in accurate energetics and other quantities \cite{knizia2013,cui2020,PhysRevResearch.2.043259,Zheng1155,PhysRevB.93.035126,PhysRevB.91.155107,Welborn2016,bulik2014}.

However, these simplifications of the DMET formulation result in a loss in ability to describe some qualitatively important physics compared to the fully frequency-dependent DMFT approach, as well as the loss of an exact self-consistency condition \cite{Tsuchimochi2015,doi:10.1063/1.5108818}.
Since the lattice description must at all times be represented by a single Slater determinant in order to obtain single-particle bath orbitals, true correlated quantum fluctuations are not able to be represented on the lattice (such as those captured by the frequency-dependence of a self-energy).
Instead, the method relies on the optimization of a static, local `correlation' potential on the lattice in an attempt to match the mean-field lattice RDM description to the correlated RDM over the fragment.
This is not always strictly possible to do, and can lead to difficulties in the numerical robustness of the method \cite{Tsuchimochi2015,doi:10.1063/1.5108818,PhysRevB.102.085123}. 
Furthermore, the bath construction for the auxiliary `cluster' model relies on projecting to the minimal bath space which exactly reproduces the static RDM over the fragment at the (static) mean-field level. However, this does not take into account the specifics of the energy-dependence of the lattice bandstructure in the coupling of the fragment to its environment in this auxiliary model, which is formally included within DMFT.

These approximations are important for the efficiency of the method, but nevertheless have physical ramifications. For instance, for a single, translationally-symmetric fragment/impurity orbital no self-consistency is possible at all, with the correlated physics found from the interacting auxiliary model unable to modify the description of the lattice. This contrasts with DMFT, where self-consistency in the explicitly dynamical self-energy can induce e.g. Kondo physics or metal to Mott insulator transitions, even at this most restrictive of fragment sizes.
For larger fragment clusters, the self-consistency in DMET still primarily relies on charge self-consistency rather than true quantum fluctuations, which are only described in the cluster, rather than the lattice description. For instance, correlation-induced phase transitions are only able to be mimicked via static symmetry-breaking via the self-consistent correlation potential. This is often a good approximation where the fragment space is large enough and static symmetry-breaking describes the physics of the phase, but can limit accuracy in other cases.

In order to overcome these limitations and rigorously reconnect to DMFT, an extension to the method was proposed, called {\it energy-weighted} density matrix embedding theory (EwDMET)\cite{edoardo2018,Fertitta2019}. This can be motivated by considering the fragment RDM as the zeroth (hole) spectral moment of the fragment Green's function, and formulating a systematic expansion by including increasing orders of these spectral moments to be self-consistently optimized (in both hole and particle sectors). In this way, the method can build increasing (implicit) resolution of the frequency-dependence of the self-consistent fragment propagator, whilst still remaining in a formally static and zero-temperature formulation. These effective, coarse-grained energetic details both describe the coupling of the fragment space to its environment (an expansion of the hybridization), as well as include a finer resolution of the different timescales of fragment quantum fluctuations induced by the local, correlated physics (an expansion of effective self-energy and propagators). 

This moment expansion spans any orthogonal polynomial representation of these dynamic quantities on the real-frequency axis to the same order (such as Chebyshev representations)\cite{PhysRevB.91.115144}. These compact representations are widely used in Green's function methods in various domains \cite{PhysRevB.98.075127,doi:10.1063/5.0003145} and have been used previously for approximate impurity solvers for the Green's function in DMFT \cite{wolf2014,PhysRevB.99.205156,lu2014,Lu2017}. However, in EwDMET there is never the need to reconstruct an explicit dynamical description at any point in the algorithm. All aspects of the self-consistency and bath space construction are similarly truncated at this order of dynamical character, and represented as static quantum variables. In this way, EwDMET can be considered a rigorous and optimal (in an orthogonal polynomial sense) truncation of the self-consistent dynamics of DMFT, rather than simply a truncation of these dynamics in the impurity solver. While no explicitly dynamic quantities are built, fully dynamical objects can be reconstructed from the converged spectral moments, directly on the real-frequency axis and without the need for analytic continuation. These can then be used to probe the correlated spectral functions, as well as derived quantities including total energies, Fermi liquid parameters and other one-particle properties, with qualitative physics such as e.g. Mott gaps emerging even at the lowest moment truncation \cite{edoardo2018}.

Furthermore, many of the efficiency and numerical advantages of DMET are retained. These primarily arise from the manifestly finite and algebraic projection to the minimal bath space of the correlated cluster model required for the chosen expansion order of the effective dynamics. This is achieved via quantum information arguments, avoiding any numerical fit of bath states \cite{PhysRevB.102.165107}, which can often be problematic for DMFT with larger clusters \cite{PhysRevB.101.035143,Nusspickelwdmft,PhysRevB.100.125165,PhysRevB.102.165107,PhysRevB.78.115102,PhysRevB.92.155126}. Furthermore, in keeping with DMET, the solution of this model still only requires numerically efficient ground-state, static quantities to construct the self-consistency. The price for inclusion of higher moment expansions for a finer implicit dynamical description, is an increase in the bath space size of the cluster model, which increases linearly with both the number of fragment orbitals, and the number of moments of the hole/particle fragment spectrum that are self-consistently optimized. This returns to the DMET bath space construction if only the zeroth and first spectral moments are desired for self-consistency,
while converging to a formally infinite bath space for all moments, in common with DMFT, where the infinite resolution of the frequency dynamics of the propagator are included. 

To ensure that these local quantum fluctuations can be self-consistently described on the original lattice, it is necessary to move beyond the DMET correlation potential \cite{knizia2012,PhysRevB.102.085123}. In order to retain a formally static approach, this is achieved in EwDMET via the introduction of a non-interacting auxiliary space which couples to the physical lattice. This auxiliary space can couple locally to the physical lattice in order to modify its band structure in response to the local fragment correlations. Each auxiliary state acts as an individual pole of an implicit self-energy, with their couplings to the physical space defining their weight. Once this auxiliary space is traced out, the physical space can exhibit arbitrary modifications of the fragment propagator, to reflect the correlation-driven changes to its spectral moments. This can induce e.g. non-idempotent lattice density matrices and opening of true Mott gaps, which are impossible with the local potential of DMET, and ensure the ability to rigorously match the desired fragment spectral moments of the lattice to those from the correlated auxiliary model.

In previous EwDMET work, the number of auxiliary states, defining the number of effective self-energy poles, was considered an arbitrary input parameter. A practical limitation of EwDMET then rapidly became the cumbersome numerical optimization of these auxiliary energies and couplings for larger impurity sizes, despite the fact that they were only ever optimized for a non-interacting model. This is similar to the numerical difficulty and ill-conditioning in the optimization of the correlation potential in DMET, where alleviating this cost has spawned alternative numerical strategies \cite{doi:10.1063/1.5108818,PhysRevB.102.085123,bulik2014,Tsuchimochi2015}. However in contrast, for DMFT this `inverse problem' of finding a self-energy which reflects the desired modifications in the fragment propagator is solved for analytically, via the Dyson~equation. Inspired by this observation, the key development in this work is to entirely remove the cumbersome and ill-conditioned numerical optimization of these auxiliary states and correlation potential, and instead directly solve for these states via an effective Dyson equation for static auxiliary states, directly within the spectral moment representation of all quantities. This drastically improves the robustness of the algorithm, and results in an overall algorithm in which no numerical fitting steps are required either for the optimization of the modification to the lattice structure (as required for DMET) or in the construction of the bath space of the auxiliary model (as required for DMFT in a Hamiltonian formulation). Furthermore, this construction rigorously defines the number of auxiliary states required for a faithful self-consistency up to a given moment order, thereby removing this technical convergence parameter from consideration in the overall method. This reformulation also ensures that the method now rapidly and rigorously converges to the dynamical mean-field theory limit, by ensuring an implicit moment expansion also of the true hybridization function.

We structure the paper as follows. In section~\ref{sec:moms}, we describe the spectral moment expansion, with Sec.~\ref{sec:algo} detailing the improved self-consistent EwDMET algorithm. The key step is the avoidance of numerical optimization of the auxiliary space considered in Sec.~\ref{sec:constructauxham}. We also describe the construction of the interacting cluster model via projection of the appropriate bath space, as well as bounds on the number of auxiliary states required for the self-consistency. We demonstrate how this reformulated bath space spans a rigorous expansion of the hybridization function, as long as the removal of local correlation is performed before the cluster projection, ensuring a rigorous and consistent truncation of the DMFT dynamics at all stages. Furthermore, connections to other methodology in the literature will be highlighted, both in the construction of an approximate environment, as well as the particularly relevant work of Lu~\textit{et~al.}, where a real-frequency DMFT scheme was proposed \cite{lu2014,Lu2017}. Section~\ref{sec:results} will present results for a number of different lattice models, focusing on the convergence with moment order for both static and dynamic quantities. We show how the optimized auxiliary space can provide the mechanism for correlation-driven quantum phase transitions, and how information such as quasiparticle weights can be directly obtained from the optimized auxiliary space. Finally, in Sec.~\ref{sec:concs} we provide a summary of the approach and prospects for the future.

\section{The Energy-weighted DMET method} 

\subsection{Moment expansion} \label{sec:moms}

The central self-consistent quantum variables of the EwDMET method are the spectral moments of the fragment propagator. These moments are separately optimized at each order~$n$ and hole/particle type, denoted as ${\bf{T}}^{(h)}[n]$ and ${\bf{T}}^{(p)}[n]$, respectively, and defined as
\begin{align}
T^{(h)}_{\alpha\beta}[n]& =\langle\Psi|{\hat c}^\dagger_\beta ~[{\hat c}_\alpha,\hat{H}]_{\{n\}}|\Psi\rangle \label{eqn:HoleMom} , \\
T^{(p)}_{\alpha\beta}[n]& =\langle\Psi| [{\hat c}_\alpha,\hat{H}]_{\{n\}} ~ {\hat c}^{\dagger}_\beta|\Psi\rangle , \label{eqn:PartMom}
\end{align} 
where $[{\hat c}_\alpha,\hat{H}]_{\{n\}}=[\dots[[c_\alpha,\hat{H}],\hat{H}],\dots \hat{H}]$ with $n$ total commutators and $[{\hat c}_\alpha,\hat{H}]_{\{0\}} = \hat{c}_\alpha$, and ${\hat c}^{(\dagger)}_{\alpha / \beta}$ denote fermionic operators acting over the considered space. These quantities are also equivalent to {\textit {energy-weighted}} one-particle reduced density matrices \cite{Fertitta2019}. More saliently, they are directly related to the local fragment propagator, as
\begin{align}
    T^{(h)}_{\alpha\beta}[n]& = \int_{-\infty}^{\mu} A_{\alpha \beta}(\omega) \, \omega^n \, \mathrm{d} \omega , \\
    T^{(p)}_{\alpha\beta}[n] &= \int_{\mu}^{\infty} A_{\alpha \beta}(\omega) \, \omega^n \, \mathrm{d} \omega ,
\end{align}
where ${\bf A}(\omega)=-\frac{1}{\pi} \im[{\bf G}^R(\omega)]$ is the (matrix-valued) spectral function derived from the retarded Green's function and $\mu$ the chemical potential. Representing the hole and particle spectral moments individually allows the low-energy dynamical structure to be resolved, while ensuring that the high-energy expansion of the central moments of the spectral distribution (corresponding to the $\frac{1}{(i\omega)^n}$ Laurent expansion of the Matsubara propagator high-frequency tail) are resolved exactly to at least order $2n+1$ \cite{Fertitta2019}.

As the number of moments increases, these moments increasingly resolve the dynamical structure of the fragment spectrum, which via the Kramers--Kronig relation also uniquely specifies the entire fragment propagator. These spectral moments defined in Eqs.~\ref{eqn:HoleMom}--\ref{eqn:PartMom} are also equivalent to the Taylor expansion coefficients of the imaginary-time Green's function expanded around the discontinuities at times $\tau=0^-$ and $\tau=0^+$, as detailed in Ref.~\onlinecite{Fertitta2019}. Finally, it should be noted that by independently including the spectral moments up to a maximum order~$\nmom$, the space of (all) polynomials of degree~$\nmom$ is spanned for each particle/hole sector, regardless of their specific form. As an example, the flexibility afforded by working with these moments allows for a reproduction of the dynamic Green's function defined by the basis of two Chebyshev polynomial expansions to the same degree, over the spectral range of the separate particle and hole states \cite{wolf2014}. In this sense of an orthogonal polynomial expansion of the real-frequency dynamics of the propagator, the optimization of these moments in EwDMET can therefore be thought of as an optimal truncation of the dynamical information in DMFT.

\subsection{An improved EwDMET formulation} \label{sec:algo}

We seek a self-consistent procedure for the optimization of the spectral moments over a arbitrary desired fragment space, up to a desired truncation, $\nmom$, which avoids all numerical optimization procedures of potentials, hybridization functions, or effective self-energies. Many of these steps are common to the previous algorithm given in Refs.~\onlinecite{edoardo2018,Fertitta2019}, where more details on some steps can be found. In Section~\ref{sec:differences}, we highlight the main changes from the previous algorithm. In the EwDMET method, there are a number of model systems which need to be defined, where the fragment is coupled to the rest of the system, bath, and/or auxiliary states. The distinction between bath and auxiliary states is primarily one of their physical effect on the spectral moments of the fragment space. Both objects represent locally fragment-coupled, non-interacting degrees of freedom, with those termed `bath' states defining the (finite-order) effect of hybridization with the rest of the system on the fragment moments, while auxiliary states are used to induce the changes of the fragment moments due to the correlated local physics (effective self-energy changes).

As with all quantum cluster methods, we require the coupled solution for these fragment moments from two models. We start with the extended `lattice' Hamiltonian $\mathcal{H}$, which defines the full system of interest and can be divided into an interacting ($\mathcal{H}_U$) and single-particle part ($\mathcal{H}_t$). 
In this work, we will be exclusively consider the single-band Hubbard model, 
    \begin{equation}
        \mathcal{H} = \underbrace{-t\sum_{<ij>,\sigma} {\hat c_{i \sigma}}^{\dagger} c_{j \sigma}}_{\mathcal{H}_t} + \underbrace{U \sum_i n_{i \uparrow} n_{i \downarrow}}_{\mathcal{H}_U},
    \end{equation}
where the lattice can be split into disjoint clusters of $n_f$ sites, where one such set of sites represents the fragment or `impurity' sites $F$, while the remaining $n_s$ translationally equivalent clusters of sites are denoted by the set $\{S\}$, with $n_f(n_s+1)$ giving the total number of sites in the lattice, $N$. 
    
This lattice model is only ever solved at a (static) mean-field level of theory. Therefore, to include the effects of the interaction-induced correlations on the fragment moments, $\naux$ non-interacting `auxiliary' states are added to the lattice Hamiltonian, representing specific poles of a local self-energy on all symmetrically-equivalent fragment clusters. We use these to define the `Weiss' Hamiltonian $\mathcal{H}_w$. In this, replicated sets of auxiliary states couple to each of these translationally equivalent fragments in $\{S\}$, with couplings $\lambda_{i \alpha}$ and energies $\varepsilon_{\alpha}$, resulting in
    \begin{eqnarray}
        \mathcal{H}_w &=& \mathcal{H}_t - \mu_{\textrm{lat}} \sum_{i \in F, \sigma}^{n_f} {\hat c}_{i \sigma}^{\dagger} {\hat c}_{i \sigma} + \sum_{s \in \{S\}}^{n_s} \sum_{i,j\in s, \sigma}^{n_f} v_{ij} {\hat c}^{\dagger}_{i \sigma} {\hat c}_{j \sigma}  \label{eq:Weissh} \\
        &+& \underbrace{\sum_{s \in \{S\}}^{n_s} \sum_{\alpha \in s, \sigma}^{\naux} \left[ \left( \sum_{i \in s}^{n_f} \lambda_{i \alpha} {\hat c}_{i \sigma}^{\dagger} {\hat c}_{\alpha \sigma} + h.c. \right) + \varepsilon_{\alpha} {\hat c}_{\alpha \sigma}^{\dagger} {\hat c}_{\alpha \sigma} \right]}_{\textrm{coupling to auxiliary system}} . \nonumber
    \end{eqnarray}
The static potential matrix~$\mathbf{v}$ with elements~$v_{ij}$, replicated over the translationally equivalent clusters, represents the static part of the self-energy and is equivalent to the `correlation potential' of DMET. Note that the auxiliary states defined by the coupling matrix ${\bf \lambda}$ and energies ${\bf \varepsilon}$ in $\mathcal{H}_w$ do \textit{not} couple to the fragment $F$, which is instead augmented only by a chemical potential, $\mu_{\textrm{lat}}$, to ensure the correct filling of electrons in the fragment space. We denote the total number of degrees of freedom in the Weiss model as $N_w=N+n_s \naux$.

\begin{figure}[h]
\includegraphics[width=0.45\textwidth]{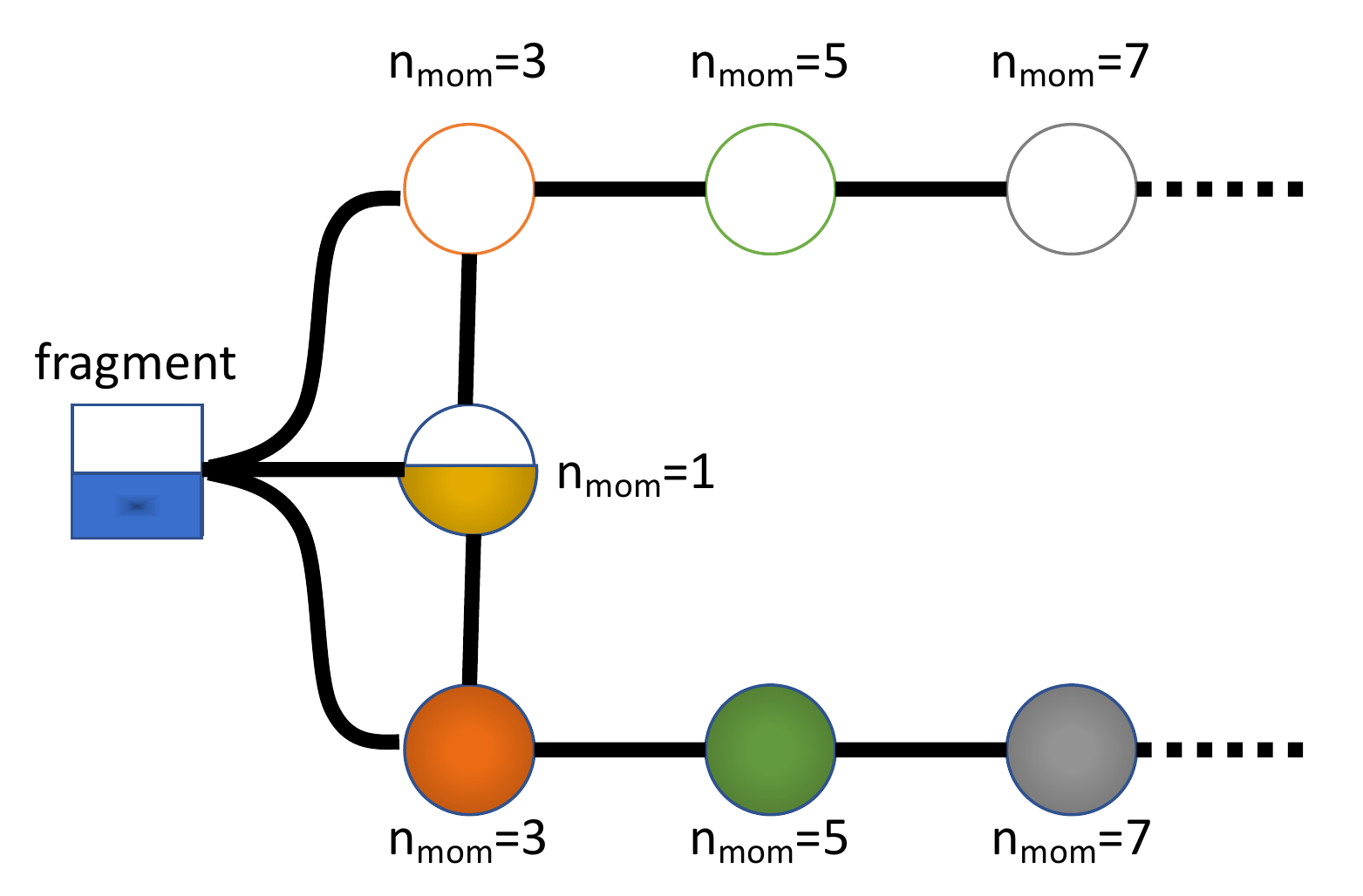}
\caption{Schematic structure of the non-interacting part of the cluster Hamiltonian ($\mathcal{H}_{cl}^{(0)}$) and bath orbital couplings in EwDMET after systematic orthogonalization. Bath orbitals are denoted by circles, with their fillings representative of the electron number when the mean-field solution of $\mathcal{H}_w$ is projected into the cluster space. Each circle represents another set of bath orbitals, whose maximum dimension is the number of orbitals in the fragment ($n_f$). Lines denote terms coupling the orbitals in the cluster Hamiltonian of Eq.~\ref{eq:clustham}. Truncating the moment expansion naturally truncates the chain of bath orbitals at a given $\nmom$.}
\label{fig:bath_visualization}
\end{figure}

This Hamiltonian~$\mathcal{H}_w$ has no explicit interaction terms, and can therefore be solved via diagonalization in the single-particle basis, leading to $\mathcal{H}_w = \sum_i^{N_w} |C_i\rangle E_i \langle C_i |$. A set of $n_b$ bath states can then be defined, which are orbitals defined across the entire support of the sites spanned by $\{S\}$ and the auxiliary states in Eq.~\ref{eq:Weissh}. The defining feature of these bath states, $|b\rangle$, is that once $\mathcal{H}_w$ is projected into the space of the fragment $\oplus$ bath (defining the `cluster' model), the resulting non-interacting spectral moments over the fragment space up to order $\nmom$ are identical to the equivalent moments resulting from the Weiss Hamiltonian model. From a DMFT perspective, these bath orbitals define the effective hybridization of the fragment to the full system for an exact reproduction of the first $\nmom$ fragment spectral moments. More details on this formal equivalence can be found in Appendix~\ref{appen:hybridexpansion}. Furthermore, this algebraic projection to the cluster model results in a manifestly finite number of bath orbitals, with an upper limit of $n_b \leq (n_f \times \nmom)$ (assuming $\nmom$ is odd, as used in this work). These bath orbitals can be constructed by orthonormalizing the set of orbitals given by
\begin{align}
    |b^{(h)}_{\alpha, m} \rangle = \sum_{\kappa \notin F}^{N_w-n_f} \sum_{E_i < \mu} E_i^m C_{\alpha i} C_{\kappa i}^* |\kappa \rangle \\
    |b^{(p)}_{\alpha, m} \rangle = \sum_{\kappa \notin F}^{N_w-n_f} \sum_{E_i > \mu} E_i^m C_{\alpha i} C_{\kappa i}^* |\kappa \rangle ,
\end{align}
where $m$ ranges from $0$ to $(\nmom-1)/2$, $\alpha$ denotes fragment orbitals, $\kappa$ denotes degrees of freedom of the Weiss Hamiltonian external to the fragment space, and $\mu$ denotes the chemical potential in the Weiss Hamiltonian required for the physical number of electrons to be correct. At least $n_f$ of these bath orbitals are linearly dependent and can be removed. The $m=0$ bath orbitals are identical to those of the DMET construction (in the absence of an auxiliary space in Eq.~\ref{eq:Weissh}). By choosing a bath representation via successive Gram-Schmidt orthogonalization of higher-order $m$ bath orbitals against lower-order ones, an interesting structure arises in the resulting cluster Hamiltonian, as visualised in Fig.~\ref{fig:bath_visualization}.
Alternatively, the bath orbitals giving rise to this structure can be directly generated by repeated singular value decomposition~(SVD) of off-diagonal blocks of the Weiss Hamiltonian.
For example, by performing a SVD on the upper block of $\mathcal{H}_w$ which couples the occupied $n_\mathrm{mom}=3$ bath orbitals to the environment external to the current cluster, the right singular vectors corresponding to non-zero singular values define the occupied $n_\mathrm{mom}=5$ bath orbitals and the remaining vectors span the orthogonal complement in the environment space.
This bath structure is identical to the `natural' bath orbital basis of Refs.~\onlinecite{lu2014,Lu2017,PhysRevB.100.115134} which had previously been highlighted as an effective bath parameterization in DMFT, though their algorithmic construction is different. Bath states of this form have also recently been highlighted as resulting in a low-entanglement solution for general impurity models, with important ramifications in their use with matrix product state methods \cite{kohn2020efficient}.
Furthermore, these bath orbitals have been generalized to capture expectation values other than spectral moments in Ref.~\cite{PhysRevB.102.165107}.

We can then define the projectors to the bath, fragment, and cluster spaces respectively as $P_b = \sum_b^{n_b} |b\rangle \langle b|$, $P_f = \sum_{i\in F}^{n_f} |i \rangle \langle i |$ and $P_c = P_f + P_b$. These are used to define an {\textit interacting} cluster model of dimensionality $n_c=n_f + n_b$, by including the interactions over the fragment space, as
\begin{equation}
    \mathcal{H}_{cl} = P_c \mathcal{H}_w P_c + P_f \mathcal{H}_U P_f - \mu_{cl} \sum_{b \notin F, \sigma}^{n_b} {\hat c}_{b \sigma}^{\dagger} {\hat c}_{b \sigma} , \label{eq:clustham}
\end{equation}
\new{where we define the single-particle part of the cluster model as $\mathcal{H}^{(0)}_{cl} = P_c \mathcal{H}_w P_c$.}
The cluster model is then solved for its ground state (in this work, exclusively via exact diagonalization), with the chemical potential $\mu_{cl}$ optimized over the bath space in order to ensure the desired number of fragment electrons, as this can change in the cluster model in the presence of interactions. The interacting spectral moments of Eqs.~\ref{eqn:HoleMom}-\ref{eqn:PartMom} can then be directly evaluated from this ground state, over all degrees of freedom in the cluster.

These resulting correlated moments can then be used to algebraically construct a new fictitious non-interacting model Hamiltonian, $h_{\textrm{aux}}$. This Hamiltonian consists of the current cluster space coupled to a set of auxiliary degrees of freedom, where the spectral moments up to order $\nmom$ over the cluster space are exactly the same as those of the interacting model of $\mathcal{H}_{cl}$ by construction. These additional auxiliary states are in a specific representation chosen such that they only couple to the cluster orbitals and not between themselves. This construction of $h_{\textrm{aux}}$ is detailed in Sec.~\ref{sec:constructauxham}. Similar to the construction of the bath space, the maximum dimensionality of the additional auxiliary space is $\naux = n_c \times \nmom$. These non-interacting auxiliary states induce changes to the spectral moments of the cluster to exactly mimic the effect of the fragment interactions.

By taking the entire cluster space, rather than just the fragment space, it is trivial to perform an effective Dyson equation on these moments, avoiding the potential for non-causal (non-hermitian) auxiliary states which can result if solely the fragment space was considered within a limited dynamics framework \cite{lu2014}. The required fragment-local auxiliary couplings and energies, as well as the static correlation potential, can be found by projecting the resulting auxiliary states into the fragment space, as
\begin{align}
    {\bf v} &= P_f \mathcal{H}_{cl}^{(0)} P_f - P_f h_{\textrm{aux}} P_f \label{eq:corrpot} ,\\
    {\bf \lambda} &= P_f h_{\textrm{aux}} (1-P_c) \label{eq:auxcoups} ,\\
    {\bf \varepsilon} &= (1-P_c) h_{\textrm{aux}} (1-P_c). \label{eq:auxe}
\end{align}
As the number of moments increases, this inversion in the entire cluster space, followed by projection into the fragment space becomes equivalent to an effective Dysons equation just within the fragment space, where the hybridization-induced components of the auxiliary states would exactly cancel bath poles and give a causal self-energy. Performing the Dyson equation in the full cluster (fragment and bath) space has previously also been employed within DMFT when working on the real-frequency axis in order to enforce causality of the resulting self-energy \cite{doi:10.1021/acs.jctc.9b00934}.
These auxiliaries represent the manifestly causal and finite number of poles of an effective fragment self-energy required to induce the correlated changes in the cluster moments, as
\begin{equation}
    {\bf \Sigma}(z) = {\bf v} + \sum_{\alpha}^{\naux} \lambda_{i\alpha} \frac{1}{z-\varepsilon_{\alpha}} \lambda_{j \alpha}^* , \label{eq:selfenergy}
\end{equation}
for an arbitrary complex frequency~$z$.

\begin{figure}[h]
\includegraphics[width=0.45\textwidth]{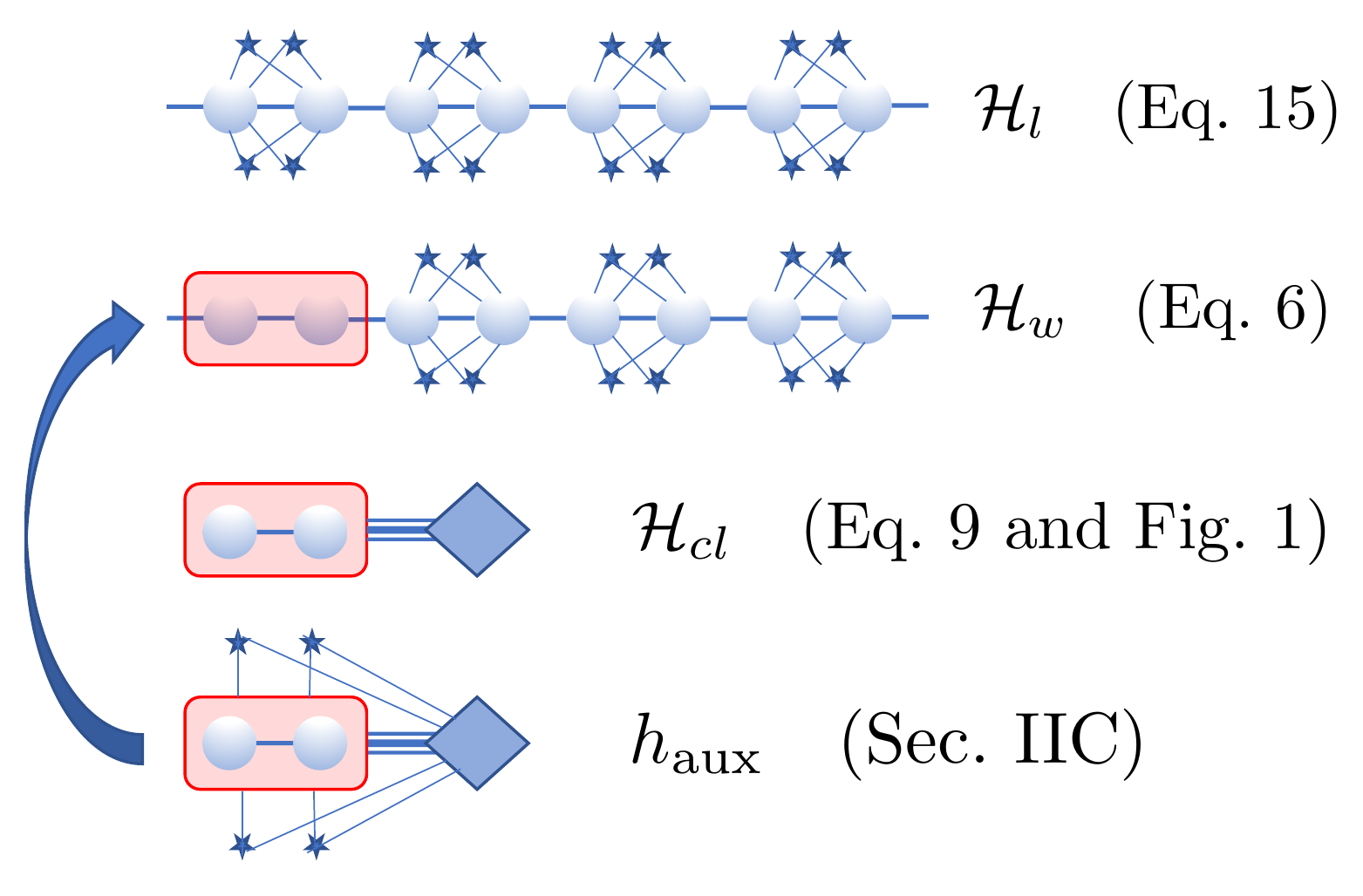}
\caption{\new{Schematic of the various Hamiltonians defined in EwDMET, for a two-site fragment (red rectangle) embedded in a one-dimensional chain of sites. Sites are denoted by circles, auxiliary states by stars, and the bath space by a diamond (where its structure and coupling to the fragment space is pictorially described in more detail in Fig.~\ref{fig:bath_visualization}). The Hamiltonians shown schematically represent the `lattice', `Weiss', `cluster' and `auxiliary' Hamiltonians from top to bottom. All Hamiltonians are non-interacting, apart from $\mathcal{H}_{cl}$, which features additional interactions over the fragment. At each iteration, $h_{\rm aux}$ is defined to have the same spectral moments as $\mathcal{H}_{cl}$.}}
\label{fig:Schematic}
\end{figure}

To complete the self-consistency, the coupling of these auxiliary states to the fragment can be used to return to the Weiss model of Eq.~\ref{eq:Weissh}, including these local auxiliary states on all symmetrically equivalent copies of the fragment, in order to update the resulting effective Weiss field. Small updates to the chemical potential $\mu_\mathrm{lat}$ as well as a constant shift in the auxiliary energies may be required to ensure the correct number of electrons in the fragment and physical lattice spaces. The algorithm can then be iterated to convergence of the auxiliary states. Expressions for local expectation values and total energies can be simply obtained from the converged fragment spectral moments of $\mathcal{H}_{cl}$ (or equivalently $h_{\textrm{aux}}$) via a representation of the Migdal--Galitskii formula in this spectral representation, as detailed in Ref.~\onlinecite{Fertitta2019}. Furthermore, explicitly frequency dependent quantities on the lattice can be reconstructed, directly from inclusion of the self-energy on all equivalent fragments, as
\begin{equation}
    {\bf G}(z) = [z\mathbf{I}-\mathcal{H}_t+\mu{\bf I}-P_f{\bf \Sigma}(z)P_f - \sum_{s\in \{S\}}^{n_s} P_s {\bf \Sigma}(z) P_s]^{-1}, \label{eq:latgf}
\end{equation}
where $P_s$ is a projector onto the space of the translationally equivalent fragment indexed by $s$. Equivalently, correlated lattice properties can be found via solution to the original Weiss hamiltonian ($\mathcal{H}_w$) with auxiliary states and correlation potential also included over the fragment space, as
    \begin{eqnarray}
        \mathcal{H}_l = \mathcal{H}_t &+& \sum_{s \in \{S\}\cup F}^{n_s+1} \left[ \sum_{\alpha \in s, \sigma}^{\naux} \left( \sum_{i \in s}^{n_f} \lambda_{i \alpha} {\hat c}_{i \sigma}^{\dagger} {\hat c}_{\alpha \sigma} + h.c. \right) \right. \nonumber \\
        &+& \left. \varepsilon_{\alpha} {\hat c}_{\alpha \sigma}^{\dagger} {\hat c}_{\alpha \sigma} - \mu \sum_{i\in s, \sigma}^{n_f} {\hat c}_{i \sigma}^{\dagger} {\hat c}_{i \sigma} \right] , \label{eq:corrlatham}
    \end{eqnarray}
which will have a spectrum equivalent to that of Eq.~\ref{eq:latgf}. The overall workflow and representations of the various Hamiltonians are schematically shown in Fig.~\ref{fig:Schematic} for a two-site fragment in a one-dimensional lattice chain.

\subsection{Construction of the auxiliary Hamiltonian}
\label{sec:constructauxham}

In this section we consider the problem of algebraically constructing a \textit{non-interacting} Hamiltonian, where auxiliary states couple to a cluster space, and modify the moments of this cluster space such that they exactly match those found from the \textit{interacting} cluster Hamiltonian of Eq.~\ref{eq:clustham}. In this form, an effective Dyson equation can be formulated in the space of spectral moments, as demonstrated in Sec.~\ref{sec:algo}, and the coupling of the resulting auxiliary states to the bath orbitals in the cluster removed (Eqs.~\ref{eq:corrpot}-\ref{eq:auxe}) in order to complete the self-consistency. This construction is inspired by the work of Lu \textit{et al.} in Ref.~\onlinecite{lu2014} where a discrete pole, single-fragment DMFT approach was formulated, but here is generalized to work with an arbitrary dimension cluster space and rigorously truncated expansion order.

The input to this step are therefore the hole and particle moments of Eqs.~\ref{eqn:HoleMom}--\ref{eqn:PartMom} over the entire cluster space of $n_c$ orbitals, which are obtained from the solution of the interacting cluster Hamiltonian of Eq.~\ref{eq:clustham}.
The first step is to split the problem into the hole and particle sectors and find two non-interacting block-tridiagonal Hamiltonians, such that the moments of these Hamiltonians return the desired moments within each sector \cite{hammerschmidt2019}. This is achieved with a modification of the block Lanczos recursion, where a series of $n_c \times n_c$ sized diagonal and off-diagonal blocks of a Hamiltonian are iteratively created, resulting in the form
\begin{equation}
      h_\mathrm{tri} = \begin{bmatrix}
   {\bf A}_0 & {\bf B}_1 & {\bf 0} & {\bf 0} & \dots \\
   {\bf B}_1^{\dagger} & {\bf A}_1 & {\bf B}_2 & {\bf 0} & \dots \\
   {\bf 0} & {\bf B}_2^{\dagger} & {\bf A}_2 & {\bf B}_3 & \dots \\
    {\bf 0} &{\bf 0} & {\bf B}_3^{\dagger} & {\bf A}_3 & \dots \\
   \vdots & \vdots & \vdots & \vdots & \ddots & 
   \end{bmatrix} . \label{eq:tridiag}
\end{equation}
This is equivalent to ensuring that the Green's function of the generated Hamiltonian is in a truncated continued fraction representation and has the same spectral moments once projected onto the cluster space (corresponding to the first block) \cite{lu2014}. The algorithm to find $\haux$ proceeds for each hole and particle set of moments (where the hole/particle labels have been dropped for brevity unless required) as follows:
\begin{enumerate}
    \item Orthonormalize the cluster space moments under the metric of the zeroth moment for each hole/particle sector, with
    \begin{equation}
        {\bf S}_n = {\bf T}[0]^{-\frac{1}{2}} {\bf T}[n] {\bf T}[0]^{-\frac{1}{2}} , 
    \end{equation}
    for $n\leq \nmom$, or ${\bf 0}$ otherwise.
    \item Build up diagonal and off-diagonal blocks of $h_\mathrm{tri}$ for each sector recursively. With ${\bf C}_0^0={\bf I}$, ${\bf A}_0={\bf S}_1$, ${\bf B}_0={\bf 0}$, and ${\bf C}_j^n = {\bf 0}$ for $j<0$, $j>n$, or $n<0$,
    these blocks can be constructed by starting at $n=0$ and iterating the following equations (in order):
    \begin{align}
{\bf B}_{n+1}^2 & = \sum_{j=0}^{n+1} \sum_{l=0}^n {\bf C}_{l}^{n\dagger} {\bf S}_{j+l+1}  {\bf C}_{j-1}^{n} - {\bf A}_n^2 - {\bf B}_n^{2\dagger } , \\
{\bf C}_j^{n+1} & = \left[ {\bf C}_{j-1}^n ~- {\bf C}_j^{n}~{\bf A}_{n} ~- {\bf C}_j^{n-1}~{\bf B}_{n}^\dagger \right] {\bf B}^{-1}_{n+1}  \label{eq:CBlock} , \\
{\bf A}_{n+1} & = \sum_{j=0}^{n+1} \sum_{l=0}^{n+1} ~{\bf C}_l^{(n+1)\dagger} ~{\bf S}_{j+l+1} ~{\bf C}_j^{n+1} .
\end{align}

\item The number of blocks in this expansion naturally truncates, since there are only $\nmom$ moments. This gives a maximum number of ${\bf A}$ blocks of $(\nmom+1)/2$, corresponding to a hard truncation of the continued fraction representation of the Green's function of the cluster. It is also possible, especially in more weakly correlated situations, that the number of generated blocks is less than this upper limit, which is evidenced when the ${\bf B}_i^2$ block becomes singular. If this happens, the final ${\bf A}_i$ block can still be computed by ensuring that only the non-null space of ${\bf B}_i$ is inverted in Eq.~\ref{eq:CBlock}, before using this to find the final ${\bf A}_i$. A Hamiltonian representation of the moments of the particle and hole space can then be found in the tridiagonal form of Eq.~\ref{eq:tridiag}.
\item The particle and hole block-tridiagonal Hamiltonians then need to be combined into a single Hamiltonian, with their combined energy spectrum and eigenvector weights in the cluster space unchanged. The individual particle and hole Hamiltonians are diagonalized as
\begin{align}
    h_\mathrm{tri}^{(h)} &= U_h E_h U_{h}^{\dagger} , \\
    h_\mathrm{tri}^{(p)} &= U_p E_p U_{p}^{\dagger} ,
\end{align}
leading to $L$ and $M$ eigenvalues and eigenvectors for the hole and particle sector, respectively. These are combined into a new $(L+M)\times(L+M)$ unitary matrix $U$, ensuring that the projection into the original cluster space for each state is maintained. The resulting system is then defined as
\begin{equation}
    h_{\textrm{comb}} = U \left[ {\begin{array}{cc}
   E_h & 0 \\
   0 & E_p \\
  \end{array} } \right] U^{\dagger} . \label{eq:combh}
\end{equation}
\item The construction of $U$ can be achieved by ensuring that the first $n_c$ components of all vectors are maintained in the combined system. We define a $n_c \times (L+M)$ matrix, $V$, where the first $L$ columns are computed as
$\frac{1}{\sqrt{2}}{\bf T}^{(h)}[0]^{\frac{1}{2}} U_{h}^{'}$ 
and the remaining $M$ columns as 
$\frac{1}{\sqrt{2}}{\bf T}^{(p)}[0]^{\frac{1}{2}} U_{p}^{'}$,
where $U_{h}^{'}$ ($U_{p}^{'}$) is a $n_c \times L$ ($n_c \times M$) submatrix of $U_h$ ($U_p$), defining the projection into the first $n_c$ rows representing the weight of the eigenstates in the cluster degrees of freedom.
The remaining $L+M-n_c$ vectors to ensure that $U$ is full-rank can be constructed from any complete, orthogonal basis which leaves the original $n_c$ vectors unchanged. We compute these as the eigenvectors corresponding to the non-zero eigenvalues of ${\bf I}_{(L+M)} - V^{\dagger} V$. These eigenvectors are added to rows of $V$, to give $U$.
\item The final non-interacting Hamiltonian, which ensures that the spectral moments projected into the first $n_c$ degrees of freedom exactly match the desired ones, is found via explicit evaluation of Eq.~\ref{eq:combh}.
\item The states in the `non-physical' space of $h_{\textrm{comb}}$ [i.e., the part external to the $n_c$ cluster orbitals, defined by the projector $P_e = (1-P_c)$] can be decoupled from each other via a diagonalization of $P_e h_{\textrm{comb}} P_e$ in order to rotate to a diagonal representation of the Hamiltonian in this auxiliary space.
This decouples these auxiliary states which induce the correlation-driven changes to the cluster spectral moments and we term this final Hamiltonian $h_{\textrm{aux}}$.
\end{enumerate}

Writing ${\bf \lambda}=P_c \haux P_e$ and $\mathrm{diag}( {\bf \varepsilon}) =P_e \haux P_e$, the resulting cluster Green's function corresponding to $\haux$ is
\begin{equation}
    {\bf G}_c(z) = \left( z{\bf I} - P_c \haux P_c - {\bf \lambda} \frac{1}{z{\bf I} - {\bf \varepsilon}} {\bf \lambda}^{\dagger} \right)^{-1} .
\end{equation}
This form is then amenable to perform Dyson equation analytically, and extract a correlation potential and auxiliary states defined by Eqs.~\ref{eq:corrpot}-\ref{eq:auxe}, with an effective self-energy to reproduce the cluster moments as given in Eq.~\ref{eq:selfenergy}.

\subsection{Differences to previous algorithm and comparison to DMFT and DMET} \label{sec:differences}

A key advantage over the previous EwDMET algorithm \cite{edoardo2018,Fertitta2019} is the formulation of an analytic approach to obtain the auxiliary states each iteration. This avoids the cumbersome non-convex numerical optimization of the previous work which rapidly became intractable for larger numbers of fragment sites. Furthermore, this work now places strict bounds on the maximum dimension of these auxiliary states, rather than considering it another technical parameter requiring convergence. 

The other substantive difference in the algorithm is that construction of the bath spaces is formulated from the Weiss Hamiltonian of Eq.~\ref{eq:Weissh}, rather than the fragment-correlated lattice Hamiltonian of Eq.~\ref{eq:corrlatham}. These resulting bath spaces are different in the presence of auxiliary states on the lattice. While the previous approach did not double count correlations, since it was the $\mathcal{H}_w$ Hamiltonian of Eq.~\ref{eq:Weissh} which was projected into the bath space, it nevertheless did not represent a faithful expansion of the true hybridization in terms of the truncated moment expansion. The non-commutativity of the operations for constructing bath orbitals, and removal of effective interactions over the fragment space (correlation potential and auxiliary states) is a subtle issue. Whether the correlation potential should be removed before or after bath space construction is also formally a consideration for DMET, where bath spaces are instead constructed in the presence of the fragment correlation potential, as was previously done for EwDMET. However, we have numerically found this not to be a significant difference for DMET in the absence of auxiliary states. It is nevertheless found to be more important for the EwDMET method, in order to ensure a rigorous and consistent truncation of the dynamics and fragment quantum fluctuations at every step. However, this does result in an inexact matching of the moments of Eq.~\ref{eq:corrlatham} and Eq.~\ref{eq:clustham} at convergence for low $\nmom$ truncation, as is true for DMFT in the limit of an inexact bath space when comparing the cluster and lattice Green's function. More details on how this updated bath construction reflects a rigorous expansion of the dynamics of the hybridization as well as the Weiss Green's function can be found in Appendix~\ref{appen:hybridexpansion}.

Comparing to DMFT, the EwDMET algorithm represents an entirely faithful reproduction of (cluster) DMFT in the limit of large $\nmom$, as is expected for a rigorous truncation of the DMFT dynamics at all steps. This formulation is however cast in an entirely zero-temperature, Hamiltonian based, and (explicitly) frequency-free formulation for all models and quantities of interest. This can enable more efficient implementations with (potentially approximate) Hamiltonian-based solvers of the correlated cluster model, as has previously been done for the full frequency dependence of DMFT \cite{doi:10.1021/acs.jctc.9b00603,PhysRevB.100.115154,PhysRevB.86.165128,PhysRevB.96.085139,PhysRevX.5.041032,wolf2014,PhysRevB.100.125165}\new{, while avoiding the ill-conditioned analytic continuation of non-Hamiltonian formulations of DMFT}. Furthermore, within the EwDMET truncation, the bath space of the cluster model is found via a projection of the full space, and is rigorously finite, with a maximum dimension which increases (linearly) with $\nmom$. This also allows us to obtain a rigorous, finite pole representation of the self-energy required for this level of description of the effective dynamics, with a number of poles which scales again linearly with both the overall cluster and fragment size. 
Some of the steps in EwDMET are influenced by the innovative work of Lu \textit{et al.}, where the DMFT equations were formulated directly in a discrete pole representation of all quantities, coupled with an approximate solver \cite{lu2014}. However, the present work avoids many of the numerical difficulties associated with the large number of poles, by systematically restricting all dynamics to this moment expansion rather than the full dynamical Green's function.

It is worth also comparing this algorithm with that of DMET. The bath space construction (and size) of DMET is identical to that of EwDMET for $\nmom=1$. \new{We also note that in DMET, two different algorithms have been used to construct the cluster Hamiltonian -- the so-called `interacting' or `non-interacting' bath formulations, depending on whether the two-body terms of the lattice are projected into the bath space or not \cite{knizia2012,bulik2014,doi:10.1021/acs.jctc.6b00316}. The EwDMET builds on the non-interacting bath formulation (similar to DMFT), since the bath space has a degree of auxiliary character. The expansion beyond the DMET bath space of $\nmom=1$} admits increasing resolution of the beyond-mean-field induced quantum fluctuations in the fragment, at the cost of a larger bath space to span the increased lengthscale of these fluctuations into the environment. Critically, the self-consistency also requires the addition of local non-interacting auxiliary states to the lattice model, rather than just a static correlation potential over the fragment space. These ensure an exact matching criteria of the desired fragment spectral moments is possible, which in this work we show can be achieved in a rigorous algebraic self-consistency, rather than a numerically ill-conditioned fit \cite{doi:10.1063/1.5108818,PhysRevB.102.085123}. Finally, we reiterate that under the reasonable assumption that the dominant computational cost of DMET comes from the solution to the interacting cluster Hamiltonian, the cost of EwDMET at $\nmom=1$ is entirely equivalent to that of DMET. In addition, the EwDMET method proposed here also provides an analytic and exact self-consistency. This self-consistency for $\nmom=1$ goes beyond just the information contained in the one-body density matrix, and also includes matching of the $n=1$ one-body energy-weighted density matrix (i.e. the Fock matrix projected into the fragment space).

\section{Results and Discussion} \label{sec:results}
\subsection{Bethe Hubbard lattice}

We first apply this updated EwDMET methodology to the half-filled Bethe Hubbard lattice. This represents an infinitely coordinated set of correlated sites in a quasi-periodic fashion. It is particularly noteworthy as a testing ground for quantum embedding methodology, as it was shown that the self-energy is entirely site-local, and therefore DMFT can converge to the exact correlated spectral properties for this system with a single fragment site \cite{PhysRevB.45.6479,PhysRevB.71.235119,PhysRevLett.83.136}. Nevertheless, the physics of the model in the paramagnetic phase is highly non-trivial, with correlations driving a phase transition from the metallic phase to a Mott insulator characterized purely by the dynamical character of the self-energy. In EwDMET, this physics needs to be captured by the specifics of the self-consistent auxiliary states in order to drive this transition. We also compare to an implementation of fully dynamical, exact diagonalization~(ED) DMFT, where seven bath orbitals are numerically fitted to the Matsubara hybridization \cite{PhysRevB.92.155126,PhysRevB.102.165107}. The numerical fitting of a limited number of bath orbitals represents the only approximation in these DMFT results to which we compare \cite{Nusspickelwdmft}.

The non-interacting density of states of the Bethe lattice is semi-circular \cite{PhysRevB.71.235119}, with a form given by
\begin{equation}
    A(\omega) = \frac{1}{2 \pi} \sqrt{4-\omega^2} ,
\end{equation}
for $|\omega|<2t$.
This was discretized by a set of 200 degrees of freedom, in order to construct the self-consistent EwDMET procedure in a computationally tractable quasi-lattice \cite{doi:10.1142/S0217979202011937,Nusspickelwdmft}. We take the odd-moment values of the maximum spectral moment, up to $\nmom=7$ (even values have the same bath size, but are self-consistent to lower-order moments than necessary, so are not considered). All cluster Hamiltonians are solved with exact diagonalization code of PySCF v1.6 \cite{https://doi.org/10.1002/wcms.1340,doi:10.1063/5.0006074}, modified to efficiently compute arbitrary orders of the spectral moments of Eqs.~\ref{eqn:HoleMom}--\ref{eqn:PartMom} via successive and direct applications of the cluster Hamiltonian after the ground state is found. Since there is only a single fragment site, the number of bath orbitals for each cluster Hamiltonian is equal to the $\nmom$ value for the calculation.

\begin{figure}[h!]
\includegraphics[width=0.5\textwidth]{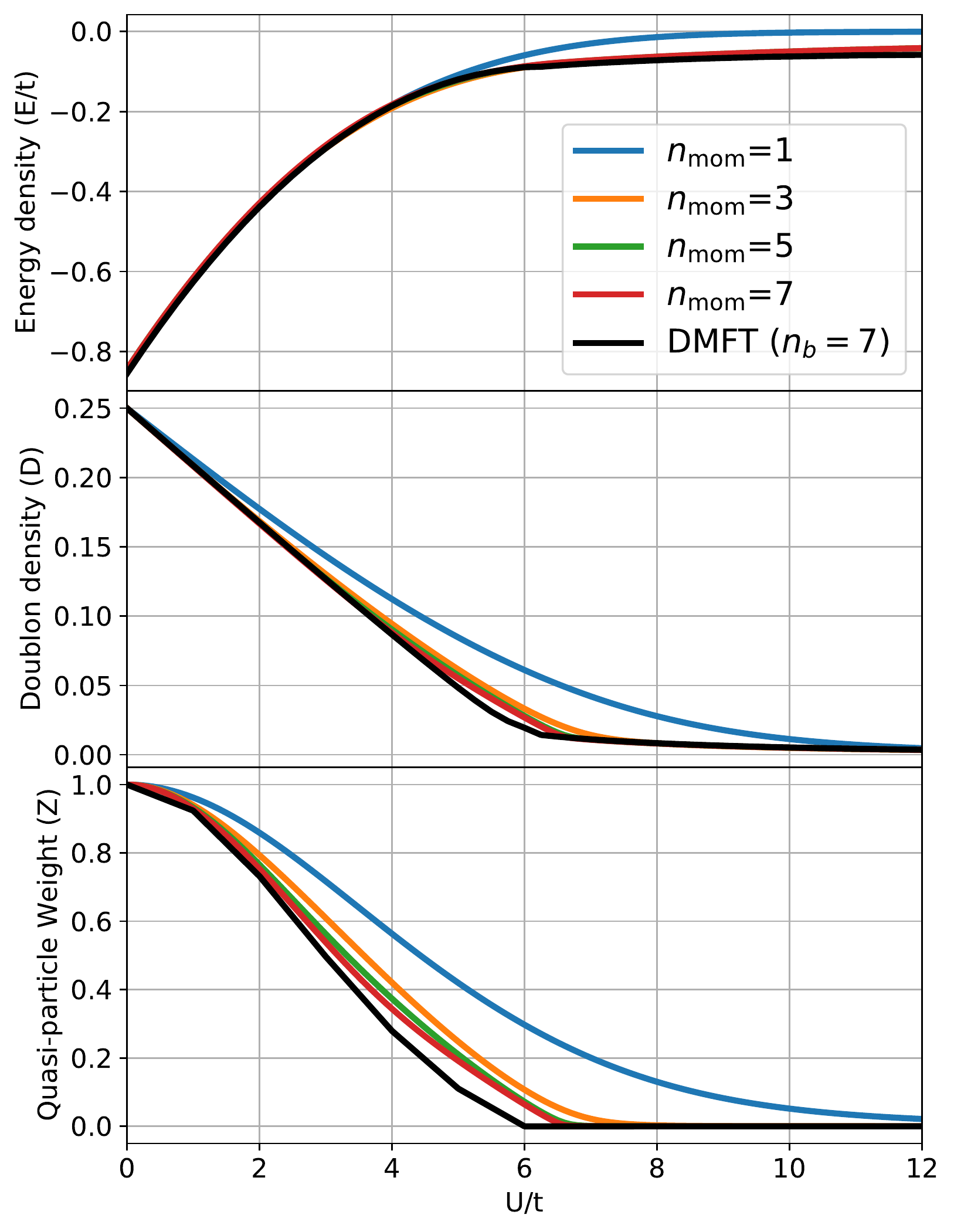}
\caption{Energy density (E/t), Doublon density (D) and quasi-particle weight (Z) for the Bethe Hubbard lattice, as the interaction strength is increased. Single fragment site EwDMET results are presented with $\nmom=1, 3, 5$, and $7$, and compared to ED-DMFT results with seven bath orbitals. 
}
\label{fig:bethe_compare_cluster}
\end{figure}

An important metric to identify and analyze the nature of any transition from metal to insulator is the quasi-particle weight of each site, which characterizes the extent to which the Fermi liquid states are renormalized by the many-body interactions. This quantity is proportional to the inverse of the effective mass of the excitations, and is defined as
\begin{align}
Z_i &= \left( 1 - \frac{\partial \Sigma(\omega)_{ii}}{\partial \omega}\Bigg|_{\omega = 0} \right)^{-1} \\
&= \left(1+\sum_{\alpha}^{\naux} \frac{|\lambda_{i \alpha}|^2}{\varepsilon_{\alpha}^2} \right)^{-1} . \label{eqn:Z}
\end{align}
This quantity is therefore directly accessible within \mbox{EwDMET} as an explicit function of the self-consistently optimized auxiliary states. Furthermore, static expectation values such as total energy density and the doublon density, defined as $\langle n_{i \uparrow} n_{i \downarrow}\rangle$, are also directly defined from the ground-state solution to the cluster Hamiltonian. Robust convergence of the EwDMET scheme was found with the auxiliary states, and these static expectation values are shown in Fig.~\ref{fig:bethe_compare_cluster} and compared to finite-bath ED-DMFT results as a function of on-site interaction strength, $U/t$. 

The energy density shows a rapid convergence with respect to $\nmom$, with all values of $\nmom\geq 3$ indistinguishable on the scale of the plot. The doublon density shows a similarly rapid convergence, albeit with a small but noticable difference between $\nmom=3$ and $5$ around the phase transition region. This phase transition is easily identified by the vanishing of the quasi-particle weight of the fragment site. As this weight decreases from one, it characterizes the increasing effective mass of the renormalized quasiparticles, with the charge becoming effectively localized in a Mott insulating state at the quantum phase transition, as $Z$ reaches zero and the effective mass diverges. For $\nmom=1$, the quasi-particle weight never quite reaches zero, characterizing an increasingly `bad' metallic state but indicating that it is unable to effectively reach the quantum critical point. However, despite the lack of a phase transition, the general trend of ``good to bad'' metal is obtained, with a single bath orbital and at an equivalent cost to DMET (which cannot describe any self-consistent dependence with $U/t$).

A phase transition is however possible for $\nmom=3$, where a slow convergence to this point is reached around $U_c \approx 8t$. As $\nmom$ is increased further, the point of the phase transition moves to lower interaction strengths down to $U_c\approx6.3t$ at $\nmom=7$ and the sharp discontinuous nature of this point is resolved. This Mott transition is a little above the finite bath ED-DMFT results, which place the transition at $U\approx 6t$, where increasing the size of the bath has previously been found to lower this point \cite{Nusspickelwdmft}. Literature values of DMFT with a fine resolution of the hybridization function and a numerical renormalization group solver place the metal to Mott insulator transition point at $U_c=5.88t$ \cite{PhysRevLett.83.136}.

\begin{figure}[h]
\includegraphics[width=8.6cm]{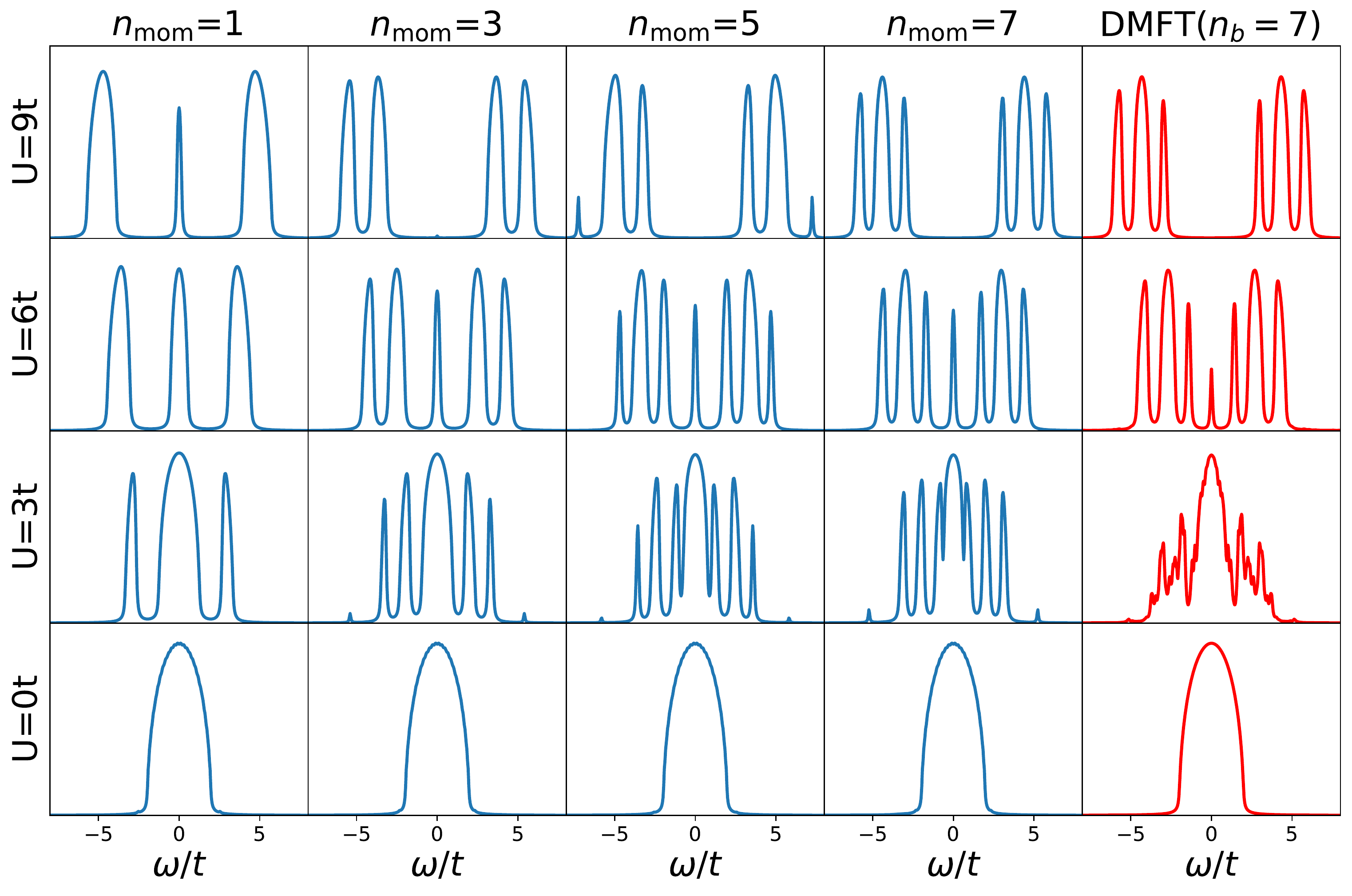}
\caption{Density of states of the Bethe Hubbard model for different values of the interaction strength $U/t$, as $\nmom$ is varied in EwDMET. Also included are comparison results from ED-DMFT with seven numerically fitted bath orbitals, which compare very favourably with the EwDMET results at $\nmom=7$. All spectra include an artificial broadening of $0.05t$.}
\label{fig:bethe_dos}
\end{figure}

We can also consider the change in the density of states resulting from the EwDMET self-consistency to different orders in the spectral moments. The density of states for different interactions are computed from the Hamiltonian of Eq.~\ref{eq:corrlatham} without requiring any analytic continuation and are shown in Fig.~\ref{fig:bethe_dos}. These show a small quasi-particle peak remaining at $\nmom=1$, with its weight decreasing with $U/t$, but never entirely disappearing, in keeping with the description of the quasi-particle weights. However, even though there is no finite-$U$ phase transition observed at this low value of $\nmom$, it nevertheless allows for the emergence of Hubbard bands in the spectrum, with these moving to higher energy as the interaction increases. The opening of a spectral gap at finite $U/t$ is seen for $\nmom \geq 3$, with additional substructure resolved in the Hubbard bands, as well as substructures in the main metallic quasiparticle peak in the Fermi-liquid phase. This structure also compares remarkably favourably, even at higher energies, with the fully dynamical DMFT with 7 bath orbitals, where in this latter approach, the full dynamics of the fragment Green's function and self-energy are explicitly and exactly solved within each iteration.

\begin{figure}[htb]
\includegraphics[width=0.5\textwidth]{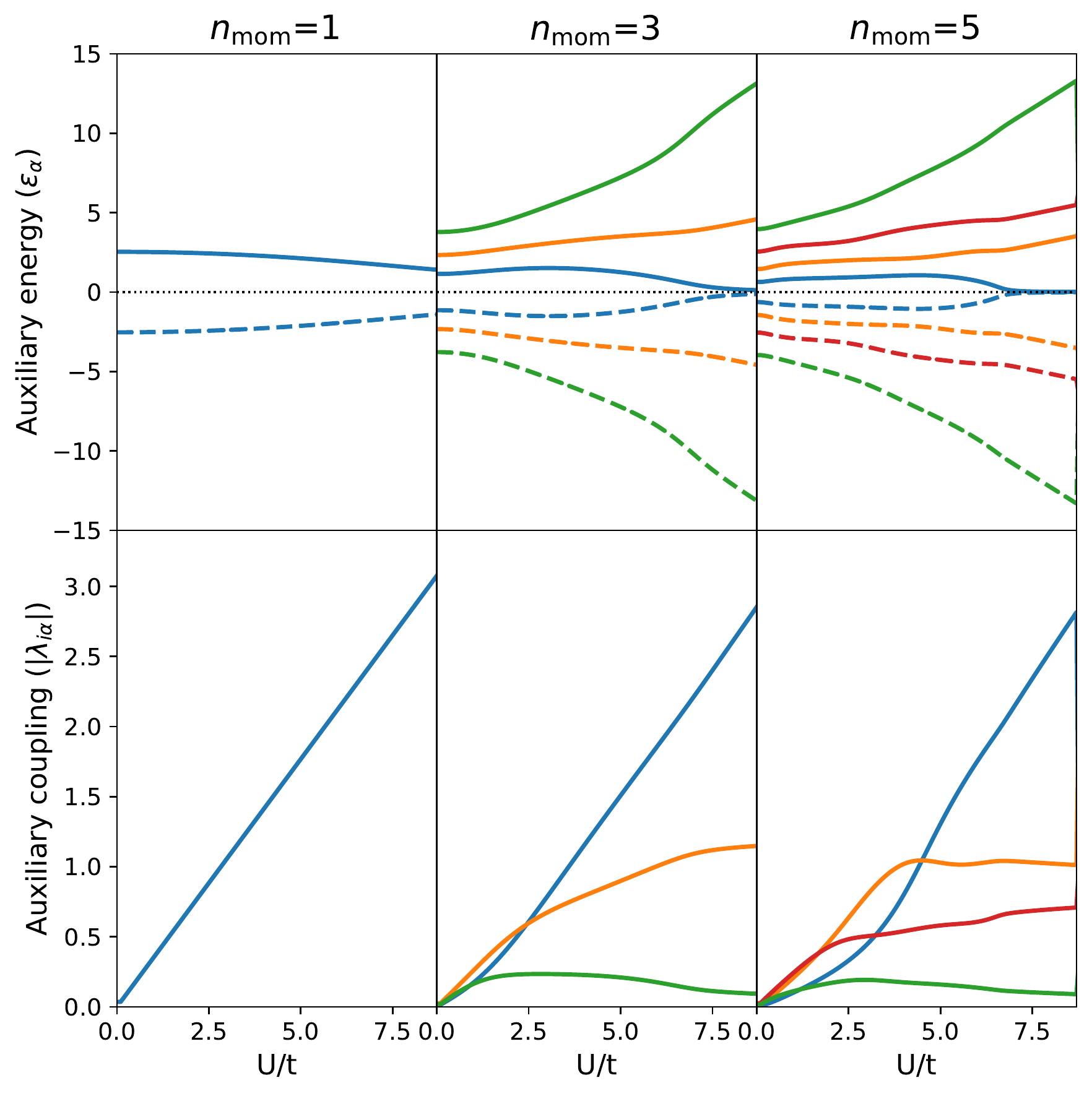}
\caption{The energies and couplings to the fragment space resulting from the optimization of the auxiliary states for different orders of self-consistency in the spectral moments, as the interaction strength $U/t$ is varied for the Bethe Hubbard model. Only auxiliary states with a coupling magnitude larger than $10^{-4}$ is shown. Results for $\nmom=7$ show a larger number of optimized auxiliary states, and is omitted for clarity. Each auxiliary state corresponds to a single pole in the effective self-energy, with Mott transitions characterized by an auxiliary state energy tending to zero\new{, while the increase in coupling to these states resulting in an opening of the Mott gap.} Auxiliary states come in particle-hole symmetric pairs \new{(denoted by dashed and solid lines of the same color)}, with equal magnitude couplings for both states in the pair. \new{Coupling and energies of the same auxiliary states are denoted by the same color.}}
\label{fig:bethe_se}
\end{figure}

We can also probe the nature of the converged auxiliary space itself, in order to understand how these states induce the phase transition, as well as describing the wider effects of the correlated physics on the implicit dynamics. Figure~\ref{fig:bethe_se} shows the optimized auxiliary energies (${\bf \varepsilon}$) and absolute values of the fragment-auxiliary couplings ($|{\bf \lambda}|$), as the interaction strength increases. Only auxiliary states with a coupling greater than $10^{-4}$ are shown. These energies represent the individual poles of an effective self-energy, with the couplings denoting the square root of the spectral weight of these poles. The auxiliary states always come in pairs with energies~$\pm \varepsilon$ and with equal couplings, reflecting the  particle--hole symmetric nature of the model. The Mott transition is triggered by an auxiliary state energy tending to zero (the chemical potential of the system), with a coupling remaining finite, as also seen as the point at which the quasi-particle weight ($Z$) tends to zero in Eq.~\ref{eqn:Z}. The gap is further widened as this zero-energy auxiliary couples to the physical site with increasing strength, which is found to scale linearly with interaction strength, as shown in Fig.~\ref{fig:bethe_se}.

\begin{figure}[htb]
\includegraphics[width=4.2cm]{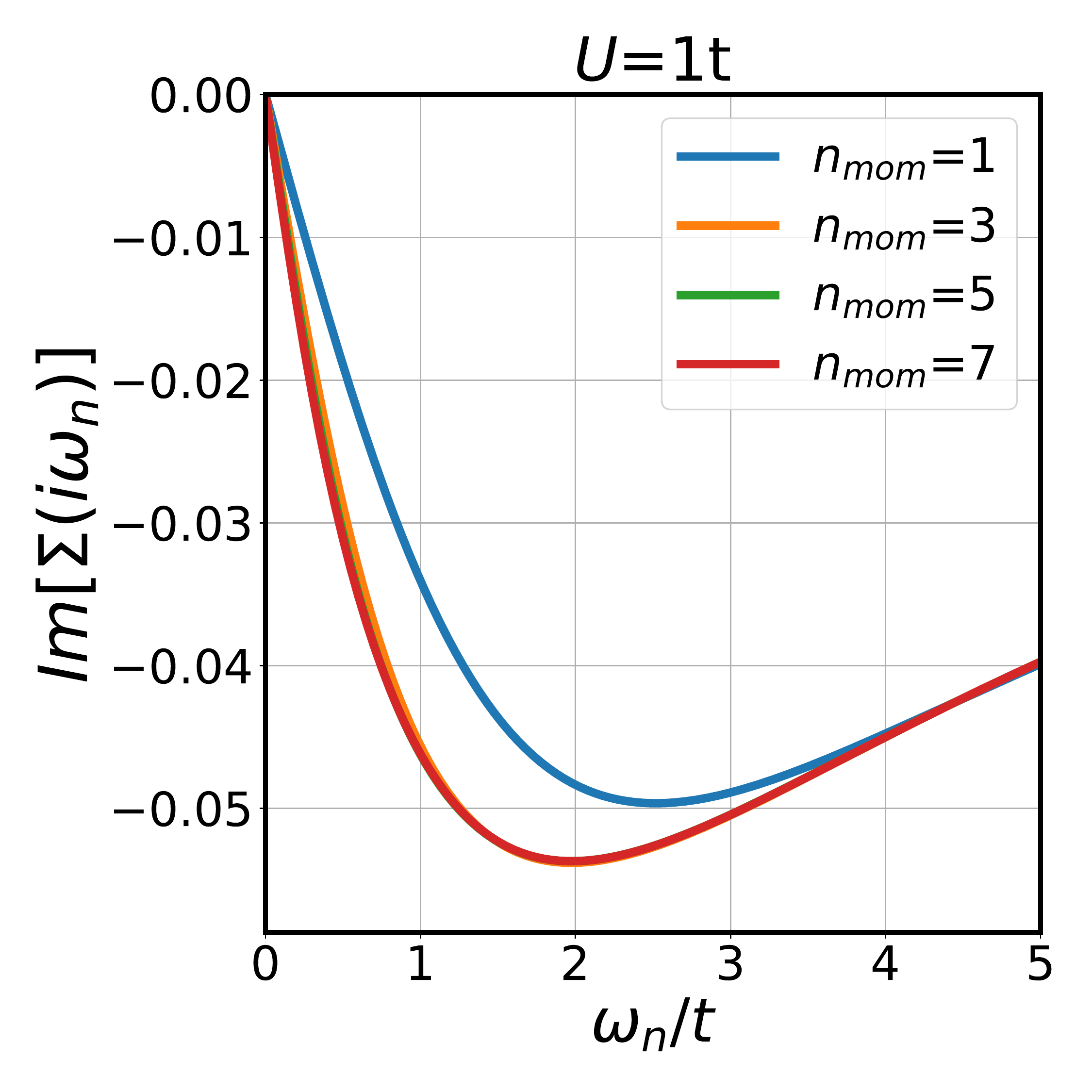}
\includegraphics[width=4.2cm]{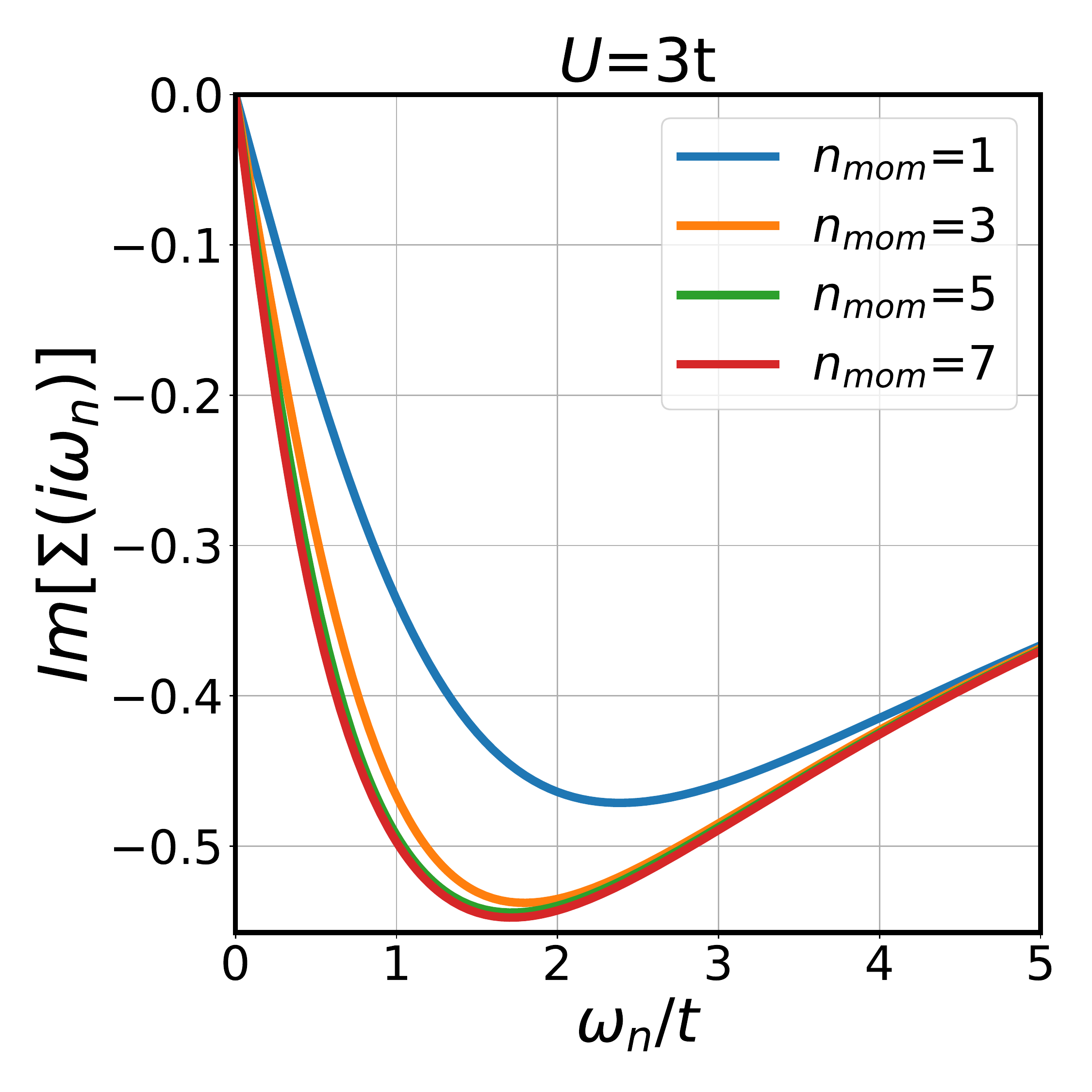}\\
\includegraphics[width=4.2cm]{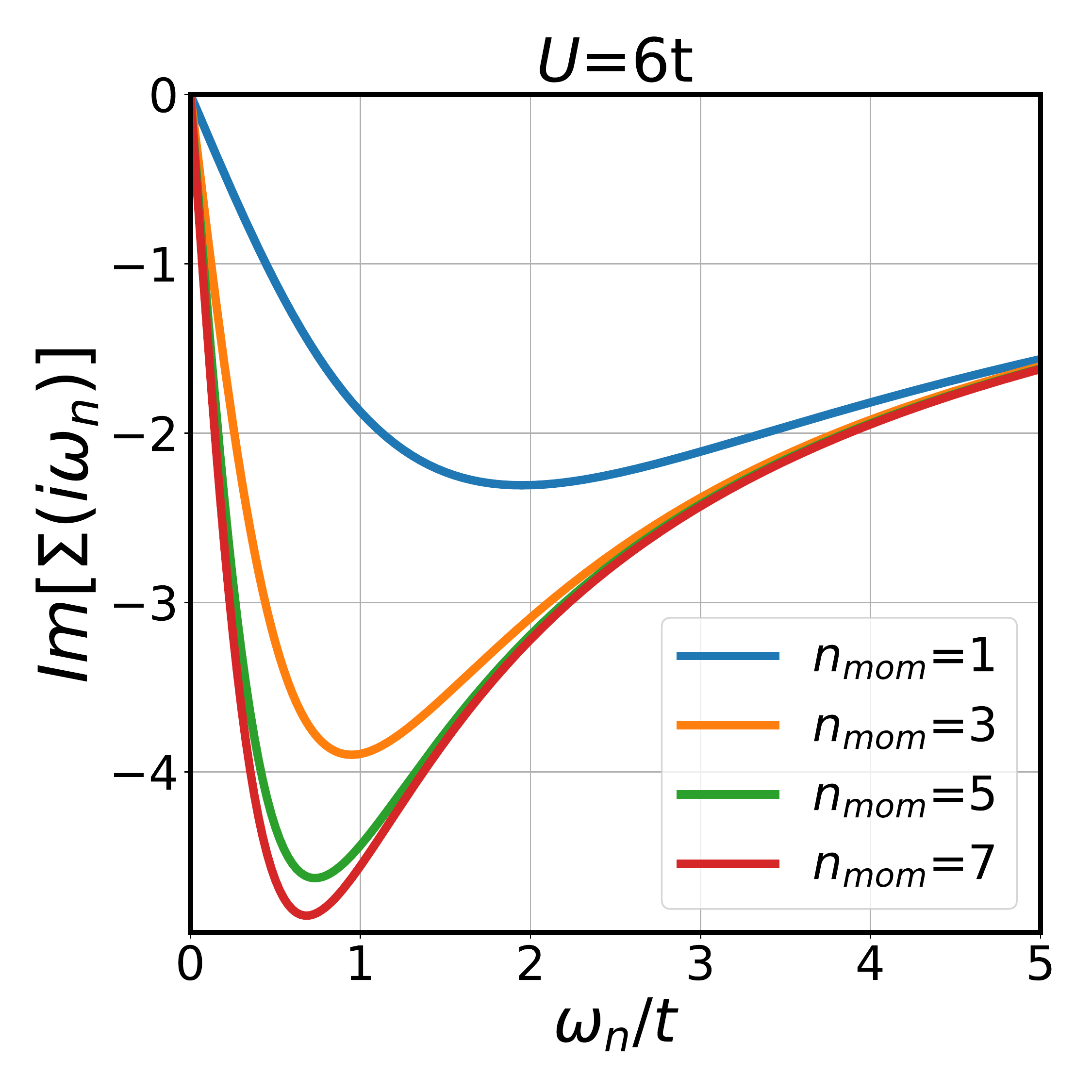}
\includegraphics[width=4.2cm]{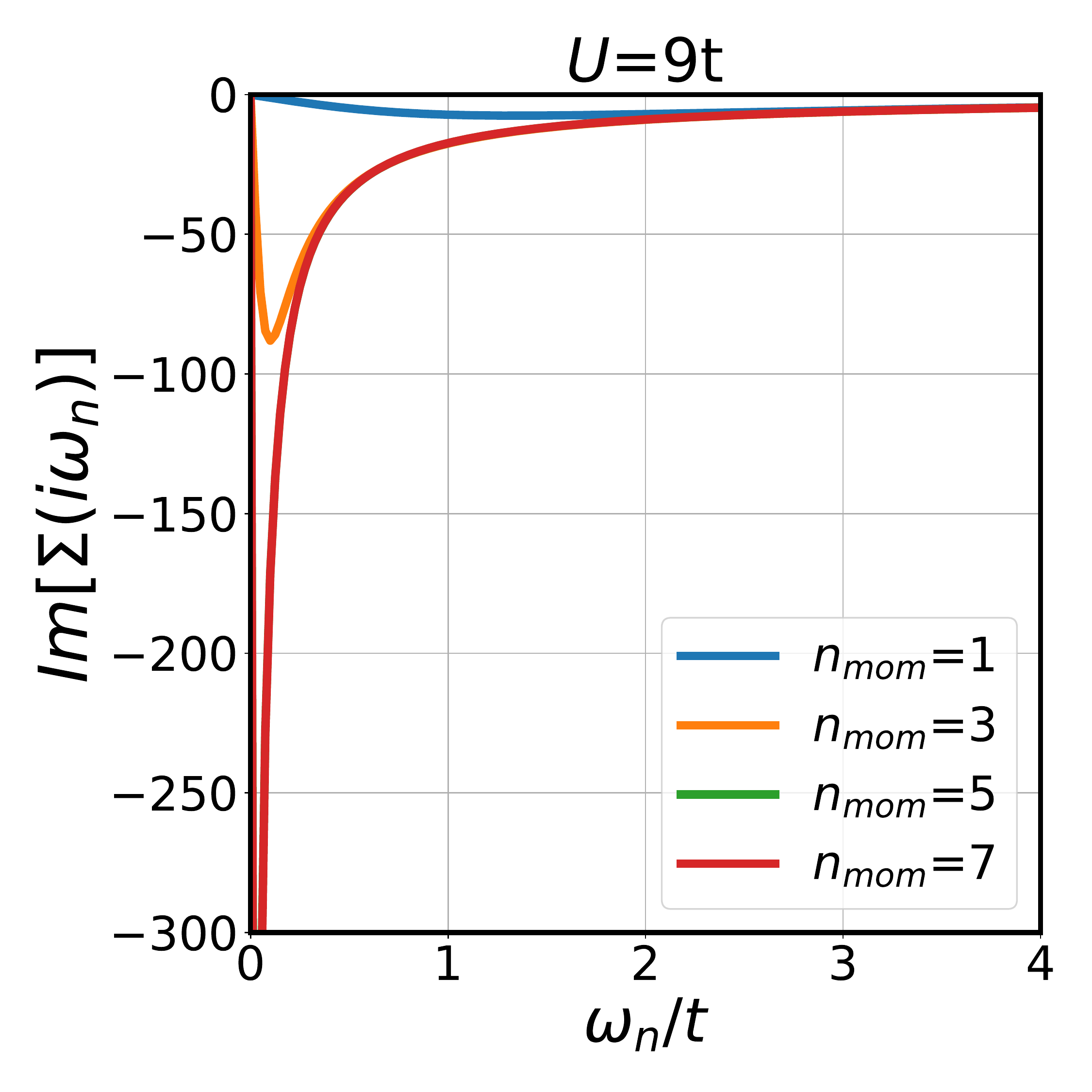}
\caption{The imaginary part of the effective self-energy on the Matsubara axis represented by the self-consistent auxiliary system for different orders of spectral moments for the Bethe lattice at $U/t = 1,3,6,9$.}
\label{fig:bethe_se_imag}
\end{figure}

Another point of note is how the number of auxiliary states changes with $\nmom$. The formal maximum number of auxiliary states is $n_c \times \nmom$, which represents a quadratic scaling with $\nmom$. However, this scaling is unphysically high due to the fact that this includes the description of correlated changes to the spectrum throughout the whole cluster (fragment and bath) space, before this is projected into just the fragment subspace afterwards. Therefore we expect this scaling to be reduced from this, especially for larger $\nmom$, as we are only interested in the fragment correlations. This is indeed observed in Fig.~\ref{fig:bethe_se}, with the number of auxiliaries tending to a more realistic linear scaling with respect to $\nmom$. This reduction is not imposed, but rather emerges naturally due to redundancies found in the auxiliary Hamiltonian construction detailed in Sec.~\ref{sec:constructauxham}, where coupling amplitudes tend to zero, and the auxiliary states can therefore be removed without approximation.
Furthermore, we can directly observe the auxiliary system convergence with $\nmom$, by considering the effective Matsubara self-energy that these auxiliary states represent via Eq.~\ref{eq:selfenergy}, as shown in Fig.~\ref{fig:bethe_se_imag}. While convergence is rapid for the Fermi-liquid phase, it is slower around the phase transition and in the Mott insulating regime, where the absolute value of the self-energy is far larger. 

Finally, it is worth noting the difference between these results, and those presented in the initial incarnation of the EwDMET method in Ref.~\onlinecite{edoardo2018}, where application to the Bethe Hubbard lattice is also reported. Section~\ref{sec:differences} details the key differences to the current approach, specifically the ability to algebraically construct the appropriate auxiliary space at each iteration, as well as the bath orbitals defined to match the moments of the Weiss Hamiltonian. These changes are found to have qualitative repercussions in the results. Previously, a gap was able to be opened at $\nmom=1$, however this was due to the fact that the number of auxiliary states was selected to be only one, which by particle-hole symmetry was constrained to be at the chemical potential, therefore forcing the opening of the gap. At higher $\nmom$, gaps were previously sometimes opened, but a clear systematic improvability of results with increasing $\nmom$ was harder to observe than with the current results. A finite interaction metal-to-insulator transition point was only found at $\nmom=4$. This is likely due to the difficulties in numerical optimization of the auxiliary states, as well as the previously unknown number required, where significant fit errors remained and a large number of possible auxiliary system solutions could be found. This numerical difficulty also precluded the extension to high moment orders with larger numbers of auxiliaries which are now possible. Overall, we find that the improved EwDMET method is able to well converge the real-frequency dynamics of the Bethe lattice system as $\nmom$ is increased, without any numerical fitting or analytic continuation, and without directly ever explicitly resolving any correlated dynamical quantity.

\subsection{One-dimensional Hubbard model}

Having considered the expansion of the effective dynamics as $\nmom$ is increased towards an exact limit in the Bethe lattice, we now consider the half-filled one-dimensional Hubbard model. For a single fragment site, this model does not result in an exact description as $\nmom$ is increased towards the complete dynamical mean-field theory, and therefore also requires an expansion in term of number of sites in the fragment ($n_f$). This is required in order to converge to a qualitatively correct description by explicitly capturing the non-local antiferromagnetic order. As the model is integrable, exact analytic results are possible in the thermodynamic limit for many quantities of interest, including the spectral gap and energy density \cite{PhysRevLett.20.1445,PhysRevLett.67.3848}. These indicate a non-zero gap opening for all values of $U/t > 0$. Furthermore, numerically exact results are possible in large lattices with the density matrix renormalization group (DMRG), which we obtain via the BLOCK v1.5.0 code \cite{doi:10.1063/1.4905329,doi:10.1063/1.3695642} in order to provide an unbiased point of comparison for static quantities.

\begin{figure}[htb]
\includegraphics[width=0.23\textwidth]{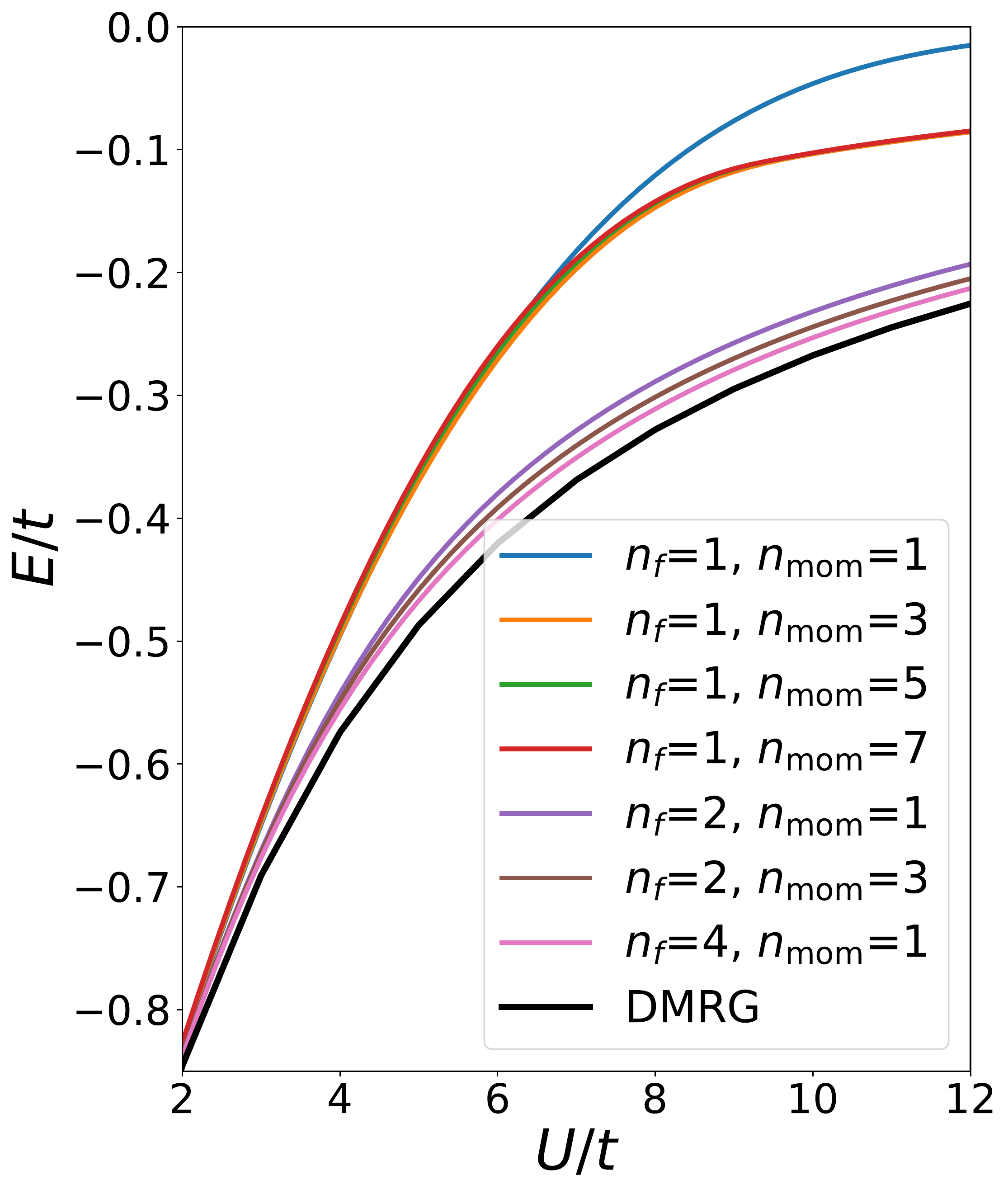}
\includegraphics[width=0.23\textwidth]{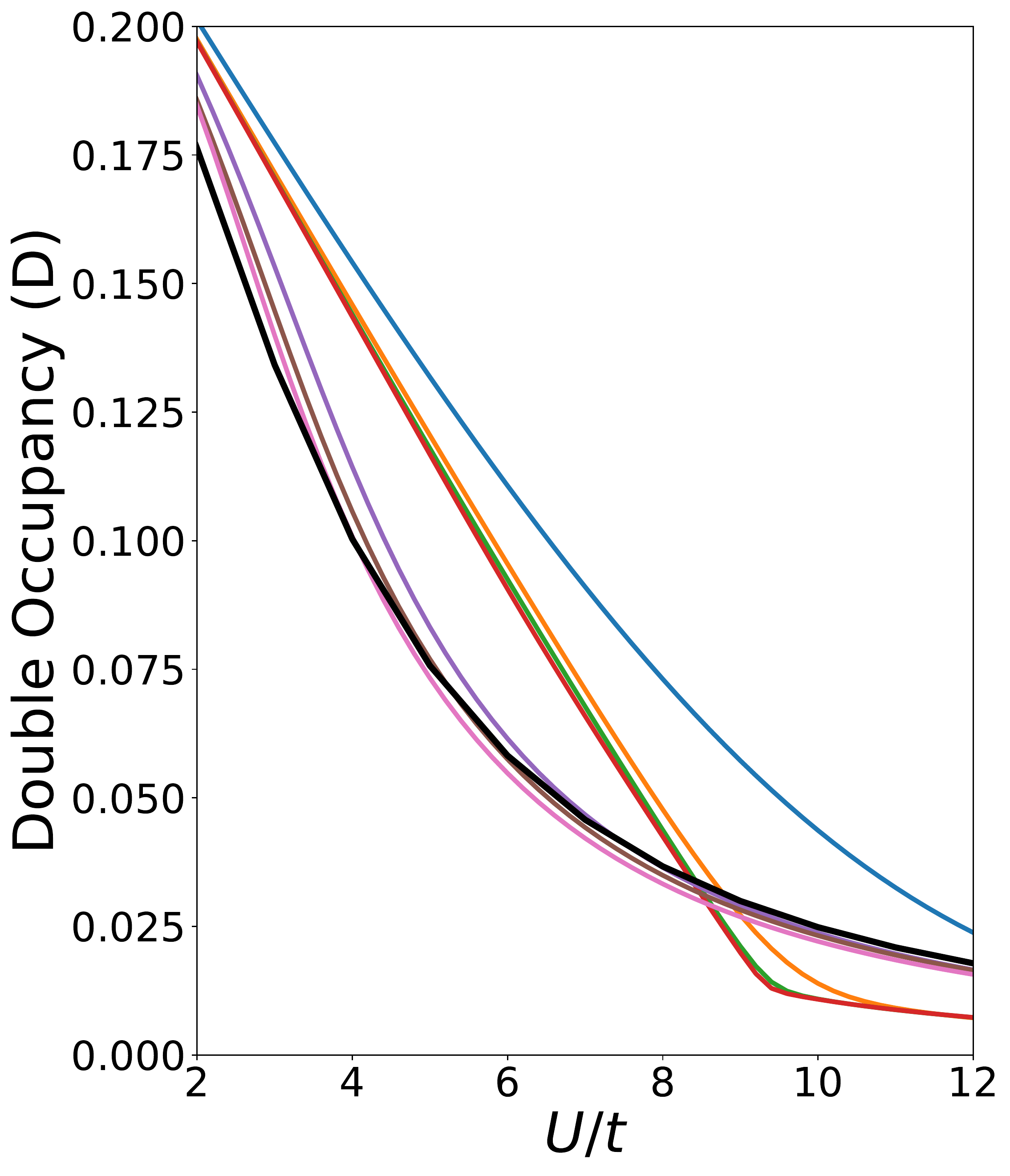}
\caption{Comparison of the energy per site and double occupancy for the 1D Hubbard chain with different numbers of fragment sites ($n_f$) and number of self-consistent moments ($\nmom$). Also included is a comparison to DMRG values.}
\label{fig:hub1d_statics}
\end{figure}

These comparisons are shown in Fig.~\ref{fig:hub1d_statics}, where expansions both in terms of the moment order and the fragment size are considered for a 144 site system with anti-periodic boundary conditions. For this system, it is found that the convergence of the effective dynamics in these quantities is very rapid. However, it is also clear that it is far more important to converge in terms of fragment size in order to get good agreement with DMRG values. The total size of the correlated cluster problem is given as $n_c = n_f(\nmom+1)$, meaning that $(n_f=1, \nmom=7)$, $(n_f=2, \nmom=3)$ and $(n_f=4, \nmom=1)$ calculations all have cluster Hamiltonians of the same size (eight degrees of freedom). It is clear that better results come from putting computational effort into increasing fragment size and therefore non-local correlated physics, rather than going to higher order in the effective dynamics and fine resolution of the longer-ranged couplings between the fragment and environment.

\begin{figure}[htb]
\includegraphics[width=8.4cm]{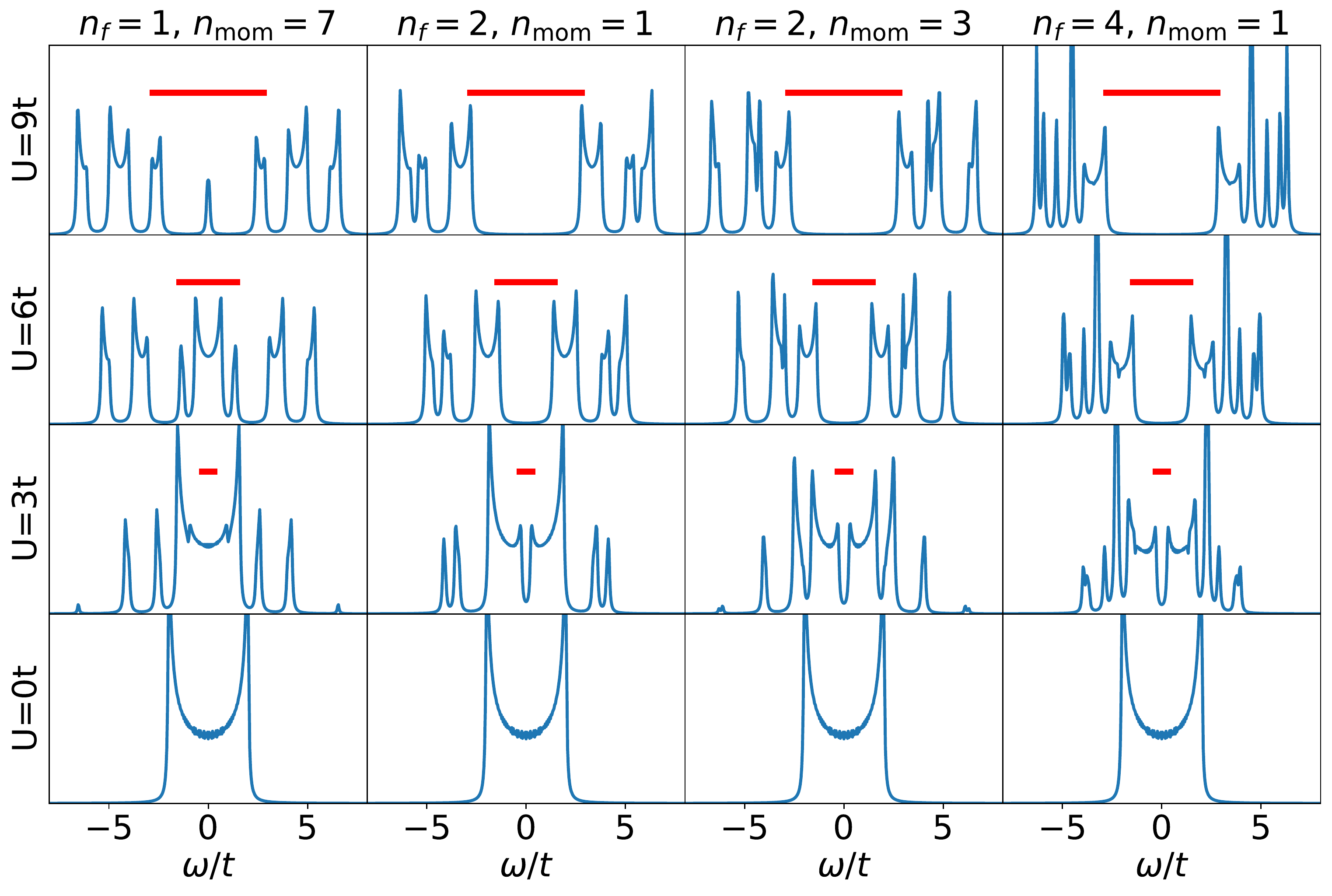}
\caption{Density of states of the 1D Hubbard chain for a selection of interaction strengths ($U/t$), fragment sizes ($n_f$) and moment expansions ($\nmom$). An artificial broadening of $0.05t$ in included for presentation purposes. Also shown as red bars are the Bethe ansatz results for the exact charge gap in the thermodynamic limit, showing improving agreement with the EwDMET results for $n_f=4$.}
\label{fig:hub1d_spectra}
\end{figure}

We can also observe the effect of fragment size and moment order on the density of states for this system, shown in Fig.~\ref{fig:hub1d_spectra}. Here, the single-fragment site results are shown to be poor, with spectral gaps opening only at very high values of $U/t$, and not at all for $\nmom=1$ (not shown). This is due to the neglect of spin symmetry breaking in the paramagnetic phase and the necessity of larger fragments for explicit antiferromagnetic order to emerge in the fragment. This mirrors DMFT results in the literature, which are similarly poor in single site approximations, even in this fully dynamical limit \cite{Go_2009}. For larger fragment clusters, the charge spectral gap is found to be quantitatively accurate compared to the Bethe ansatze results, as shown in Fig.~\ref{fig:hub1d_spectra}. Once again, increasing $\nmom$ further shows only modest changes in the spectrum, with the gap and low energy physics well converged at $\nmom=1$. This relative unimportance of the details of the propagator dynamics in this model has been observed previously \cite{edoardo2018,Nusspickelwdmft,Go_2009}.


\subsection{Two-dimensional square Hubbard model}

The 2D square Hubbard model is a particularly challenging model of significant interest due to its anticipated mapping from a number of strongly correlated materials of interest, including the parent compounds of cuprate superconductors \cite{macridin2005}. Explicit correlations within a $2\times 2$ plaquette of sites are crucial to consider, as these define the minimal unit cell of many of the ordered states expected to dominate the physics of the system (although many are also much larger \cite{Zheng1155}). This places severe restrictions on the ability to solve the cluster model with reasonable bath sizes. In the paramagnetic phase considered here, the lack of ability for the environment to generate static long-range magnetic order ensures that the model has a finite-$U$ Mott transition driven by the fragment correlations, which is expected to be representative of many paramagnetic Mott-insulator transitions in correlated materials \cite{PhysRevLett.86.5345,PhysRevB.91.125109,PhysRevLett.101.186403}.

\begin{figure}[h]
\includegraphics[width=0.45\textwidth]{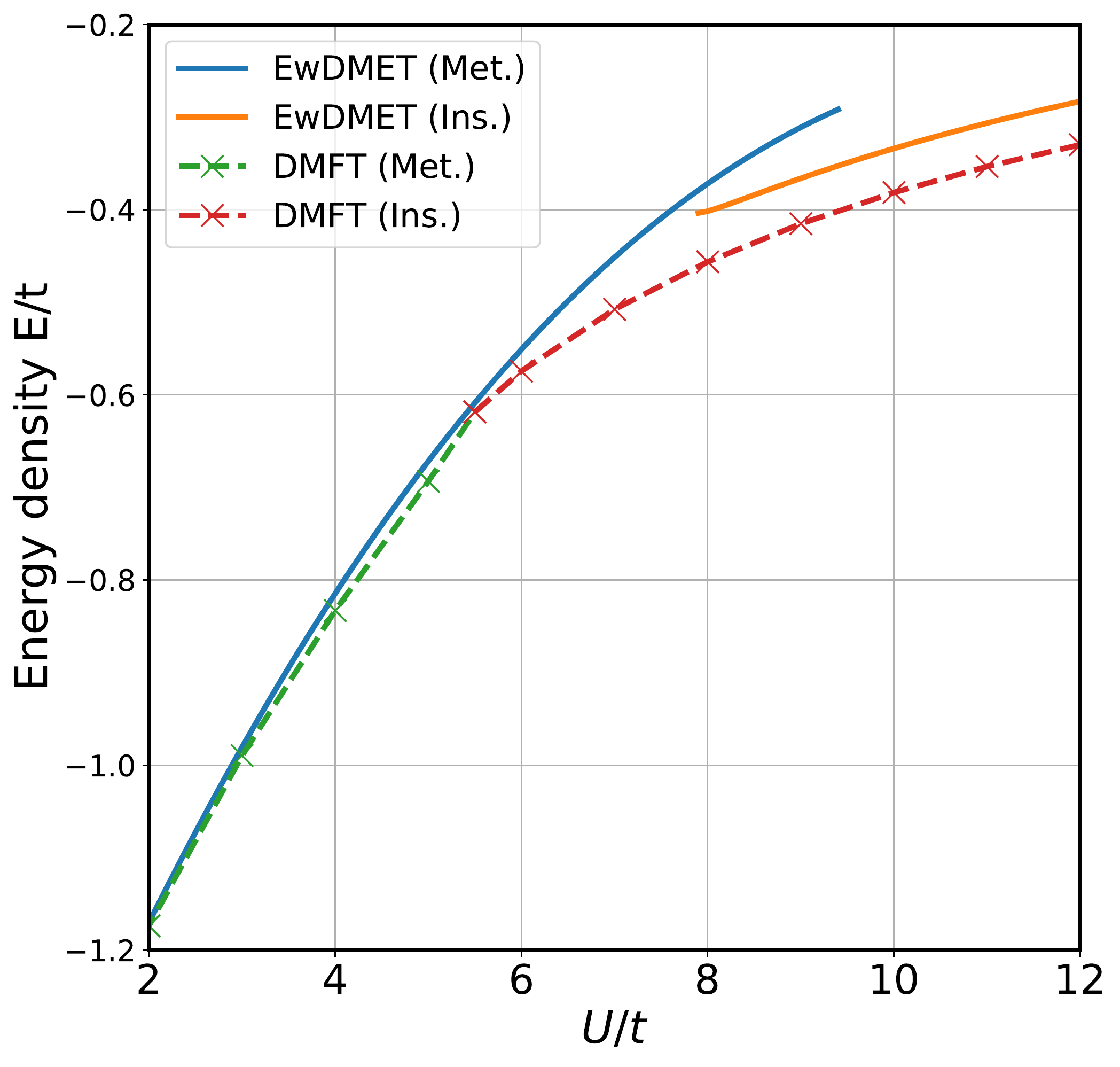}
\caption{The energy per site for the 2D Hubbard model on a $48 \times 24$ site lattice, via EwDMET and DMFT with a $2 \times 2$ fragment plaquette. Results are distinguished by the phase of the converged solution found (metallic or insulating). EwDMET results are converged for $\nmom=1$, with four bath orbitals. Comparison DMFT results are found via exact diagonalization and a numerical fit of the Matsubara hybridization at each iteration to eight bath orbitals.}
\label{fig:hub2d_eps}
\end{figure}

Figure~\ref{fig:hub2d_eps} shows the energy per site for the half-filled 2D Hubbard model on a $48 \times 24$ site lattice, a $2\times2$ fragment plaquette, and $\nmom=1$ (four bath orbitals). The results show the transition from a metallic phase at lower interacting strengths, to a paramagnetic Mott insulating phase. For comparison, we also include DMFT results with eight numerically fit bath orbitals. The first-order transition point is found to be at $U_c \approx 7.5t$, while the ED-DMFT transition is lower, at $U_c \approx 5.5t$. It is anticipated that enlarging the bath size of DMFT further will push the transition point to higher interaction strengths (towards EwDMET), since using only six bath orbitals gives a lower DMFT transition of $U_c \approx 5.0t$, while the $\omega$-DMFT approach, where the DMFT bath space is dynamically adapted for each frequency point, gives a higher transition point of $U_c \approx 6.25t$ \cite{Nusspickelwdmft}. Estimates for this transition point and plaquette size can also be found from extrapolation of finite-temperature DMFT with QMC solvers to zero temperature, yielding $U_c\approx5.8t$ \cite{PhysRevLett.101.186403}. This suggests that increasing the dynamical character of EwDMET would indeed lower the transition point, but not as much as the DMFT results of Fig.~\ref{fig:hub2d_eps} might suggest. In contrast, the DMET results in Fig.~5 of Ref.~\onlinecite{knizia2012} suggest a paramagnetic metal to insulator crossover point at a higher interaction strength compared to all methods, with $U_c \approx 9.5t$. Furthermore, the DMET also predict an insulating metastable solution at lower interactions than $U_c$, which is not found for either EwDMET or DMFT results.

\new{Finally, we note that the convergence of the algorithm for all interaction strengths is found to be robust, with the majority of points reliably converging in $\sim10$ iterations. However, close to the phase transition, where there are competing low-energy solutions, we note that the number of iterations substantially increases, to a maximum of 130 iterations overall (a similar pattern was also observed for the $\nmom=7$ Bethe lattice results). We anticipate that convergence acceleration schemes such as DIIS\cite{BANERJEE201631}, or even simple damped updates to the auxiliary parameters would improve this convergence speed, as we are currently using a simple undamped forward iteration algorithm. These will be investigated in the future, as ways to accelerate convergence. It should also be noted that similar to DMET and DMFT, the optimization parameters (in this case, the auxiliary energies, couplings and correlation potential), can also be initialized from prior converged values at `nearby' interaction strengths. This can improve convergence speed, but more importantly allow for the following of metastable solutions, as shown for the metallic state of Fig.~\ref{fig:hub2d_eps}.}

\section{Conclusion} \label{sec:concs}

In this work we have presented a reformulation of the energy-weighted density matrix embedding theory (EwDMET). This method is now cast rigorously as both a way to systematically improve the DMET method via a self-consistent capturing of increasingly long-range quantum fluctuations out of the fragment space, or as a way to systematically and rigorously truncate the explicit dynamics of dynamical mean-field theory. The approach is formulated as a ground state, zero-temperature method, and avoids any necessity for numerically challenging fitting of bath orbitals, a correlation potential, or auxiliary states, as found in Hamiltonian formulations of DMFT, DMET, and the original presentation of the EwDMET method, respectively. The approach therefore combines many of the strengths of these parent methods, with a rigorous and algebraic self-consistency which can be fulfilled exactly.

We extensively benchmark this new EwDMET formulation against a number of Hubbard models in different physical domains. For the infinitely coordinated Bethe lattice we show how longer ranged quantum fluctuations / implicit higher-order dynamics of the fragment propagator are essential in order to obtain a phase transition, and observe the ability to converge even the higher-energy dynamics of the spectrum to excellent agreement with the exact limit as given by dynamical mean-field theory. Quantities such as quasiparticle renormalization and other Fermi liquid parameters can be analytically extracted from the converged auxiliary space directly at zero temperature and without fitting or analytic continuation. In the one-dimensional model, the effect of increasing the dynamical resolution is small, allowing EwDMFT to focus efforts on enlarging the fragment size. This is achieved for fragments including up to four sites, while truncating the dynamics to only $\nmom=1$ and ensuring a bath space which is rigorously the same size as the fragment space. These results give excellent agreement for both the statics compared to DMRG results, and the charge gap compared to analytic values in the thermodynamic limit. Finally, a $2\times2$ site fragment plaquette is used for the square 2D Hubbard lattice to find a paramagnetic metal to Mott-insulator transition. The found transition point is between the fully dynamical DMFT result estimated with a finite bath fitting or finite temperature extrapolation and previously reported DMET results, indicating that increasing the self-consistent dynamical resolution will likely lower the transition point. Nevertheless, even with limited dynamical content, the results in this challenging case are in good agreement and benefit from the compact and analytic bath construction and self-consistency.

The ability to remove all fitting steps or analytic continuation from this quantum embedding, whilst still working in a zero-temperature formalism, suggests that it is an ideal candidate for use with significantly larger fragment sizes, where fitting difficulties become more serious with alternative approaches. Future work will aim to move the approach beyond benchmarking, into larger systems and fragment sizes with a range of Hamiltonian-based solvers, symmetry-broken phases, \new{doping}, and coupling to long-range interacting effects, thereby extending the scope of this approach beyond the particle--hole symmetric, local Hamiltonians considered here.

\begin{acknowledgements} 
G.H.B. gratefully acknowledges support from the Royal Society via a University Research Fellowship, as well as funding from the European Research Council (ERC) under the European Union’s Horizon 2020 research and innovation programme (grant agreement No. 759063). We are also grateful to the UK Materials and Molecular Modelling Hub for computational resources, which is partially funded by EPSRC (EP/P020194/1 and EP/T022213/1).
\end{acknowledgements}

\appendix
\section{The EwDMET bath space as a formal hybridization expansion}
\label{appen:hybridexpansion}


In the earlier work of Refs.~\citenum{edoardo2018,Fertitta2019} the Hamiltonian
of Eq.~\eqref{eq:corrlatham}, which includes auxiliary states representing the effect of the local self-energy on the fragment itself, was used to construct the bath orbitals,
rather than the Hamiltonian of Eq.~\eqref{eq:Weissh} used in this work.
Previously, these local auxiliary degrees of freedom were then projected out of the bath orbitals {\it after} their construction (rather than before, which gives the Weiss Hamiltonian), in order to avoid a double counting of the local correlation effects.
While this ``diagonalize--then--project'' scheme ensures reproduction of the moments of the correlated lattice, rather that Weiss Hamiltonian, it also leads to
non-normalized bath orbitals, as weight on the bath orbitals remains on uncoupled degrees of freedom.

\begin{figure}[h]
\includegraphics[width=1.0\linewidth]{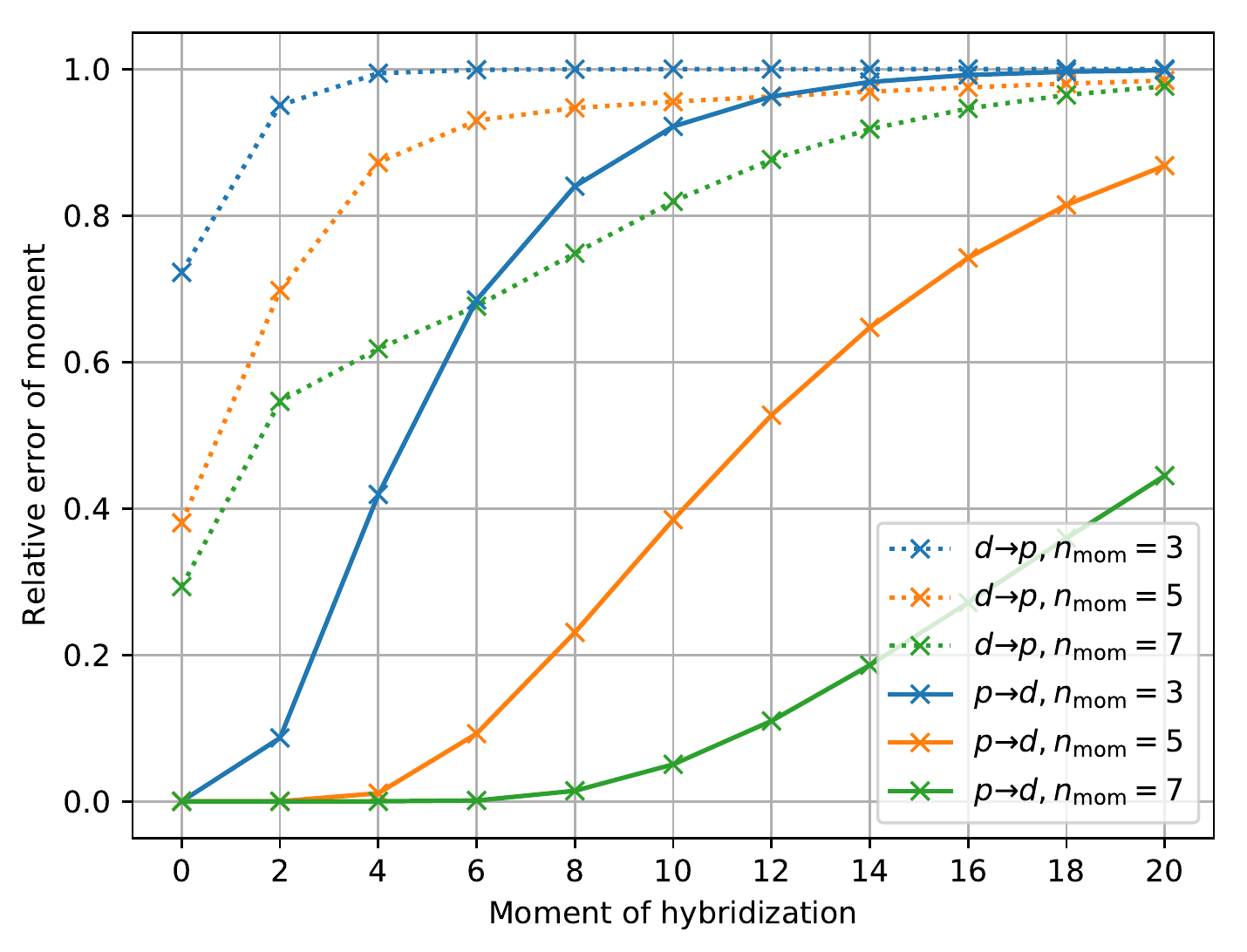}
\caption{
Relative errors in the moments of the hybridization comparing
the ``diagonalize--then--project'' ($d\to p$) scheme used in previous publications and the
``project--then--diagonalize'' ($p \to d$) scheme used in this work.
The $d \to p$ approach does not lead to an exact matching of hybridization moments.
The $p \to d$ method leads to an exact matching of $n_\mathrm{mom}$ moments of the Green's~function and $n_\mathrm{mom}-2$ moments of the hybridization. 
Only the even moments are shown, since all odd moments of the hybridization are zero for the half-filled one-dimensional Hubbard-model.
}
\label{fig:compare_bath_construct}
\end{figure}

In this work an alternative ``project--then--diagonalize'' approach is taken, which is now consistent with the approach of DMFT. In this, the local auxiliary states are never added to the Weiss Hamiltonian over the fragment space, and this is the self-consistent lattice Hamiltonian from which bath orbitals are constructed.
In contrast to the previous method, the resulting bath orbitals are normalized and
result in a systematic matching of moments of the Green's~function {\it and} hybridization. This is shown in Fig.~\ref{fig:compare_bath_construct} for the example of the
one-dimensional Hubbard model, where two auxiliaries at energies~$\pm 10^{-6} t$
and with coupling strength~$0.1 t$ were added for each fragment consisting of two lattice sites.
In addition to guaranteeing the exact matching of $n_\mathrm{mom}$~moments of the Weiss Green's~function and $n_\mathrm{mom}-2$~moments of the hybridization,
the bath orbitals of the ``project--then--diagonalize'' approach also result in lower errors for the higher moments in an hybridization expansion.

%
%


%

\end{document}